\documentclass[aps,twocolumn,prb,10pt, floatfix,longbibliography,final]{revtex4-2}
\usepackage[caption=false]{subfig}
\usepackage{graphicx, amsmath, bm, amsfonts, amssymb, color}
\usepackage[%
  pdftex,
  breaklinks=true,
  colorlinks=true,
  linkcolor=blue,
  urlcolor=blue,
  citecolor=blue
]{hyperref}
\usepackage{verbatim}
\usepackage{xcolor}
\graphicspath{ {./images/} }

\newcommand{\beq}[0]{\begin{equation}}
\newcommand{\eeq}[0]{\end{equation}}

\usepackage{enumitem}

\newcommand{\Ch}[1]{\color{black}#1\color{black}}

\newcommand{\Mat}[1]{\bm{{\sf #1}}}

\begin{document}

\title{\Ch{Flux noise in disordered spin systems}}
\author{Jos\'{e} Alberto Nava Aquino}
\author{Rog\'{e}rio de Sousa}
\affiliation{Department of Physics and Astronomy, University of Victoria, Victoria, British Columbia V8W 2Y2, Canada}
\affiliation{Centre for Advanced Materials and Related Technology, University of Victoria, Victoria, British Columbia V8W 2Y2, Canada}

\date{\today}

\begin{abstract}
Impurity spins randomly distributed at the surfaces and interfaces of superconducting wires are known to cause flux noise in Superconducting Quantum Interference Devices (SQUIDs), providing a dominant mechanism for decoherence in flux-tunable superconducting qubits. 
While flux noise is well characterized experimentally, the microscopic model underlying spin dynamics remains a great puzzle.
\Ch{The main problem is that first-principles theories based on integration of the quantum Heisenberg equations of motion for interacting spins are too computationally expensive to capture spin diffusion over large length scales, hindering comparisons between microscopic models and experimental data. 
In contrast, third principles approaches lump spin dynamics into a single phenomenological spin-diffusion operator $D\nabla^{2}$, that is not able to describe the quantum noise regime and connect to microscopic models and different disorder scenarios such as spin clusters}. 
Here we propose an intermediate ``second principles'' method to describe general spin dissipation and flux noise in the quantum regime.
It leads to the interpretation that flux noise arises from the density of paramagnon excitations at the edge of the superconducting wire, 
with paramagnon-paramagnon interactions leading to spin diffusion, and 
interactions between paramagnons and other degrees of freedom such as phonons, electrons, and two-level systems leading to spin energy relaxation. 
At high frequency $\omega$ we obtain an upper bound for flux noise, showing that the (super)Ohmic noise observed in experiments is not originating from interacting spin impurities. 
We apply the method to Heisenberg models in two dimensional square lattices with a random distribution of vacancies, with nearest-neighbor spins coupled by constant exchange. 
Explicit numerical calculations of flux noise show that it follows the observed power law $A/\omega^{\alpha}$, with amplitude $A$ and exponent $\alpha$ depending on temperature and inhomogeneities such as spatial confinement and disorder. These results are compared to experiments in niobium and aluminum devices.
The method  establishes a connection between flux noise experiments and microscopic Hamiltonians with the goal of identifying relevant microscopic mechanisms and guiding strategies for reducing flux noise. 
\end{abstract}
\maketitle

\section{Introduction\label{sec:introduction}}

While progress in experimental realization of quantum computers based on superconducting wires and Josephson junctions has been remarkable \cite{Arute2019}, the noise level in current devices greatly reduces their capacity to solve problems, and washes out their  ``quantum advantage''. One key issue is the trade off between scalability and flux noise. 
Qubit frequency tunability is essential to circumvent the frequency crowding problem faced by superconducting circuits with more than 100 qubits. This requires the addition of Superconducting Quantum Interference Devices (SQUIDs) to the circuit, increasing their sensitivity to flux noise \cite{Hutchings2017, Chavez-Garcia2022}. 
A similar issue plagues SQUID qubits, in that additional qubit interconnection increases the impact of flux noise \cite{Zaborniak2021}. 

\begin{figure}
	\centering
	\includegraphics[width=0.5\textwidth]{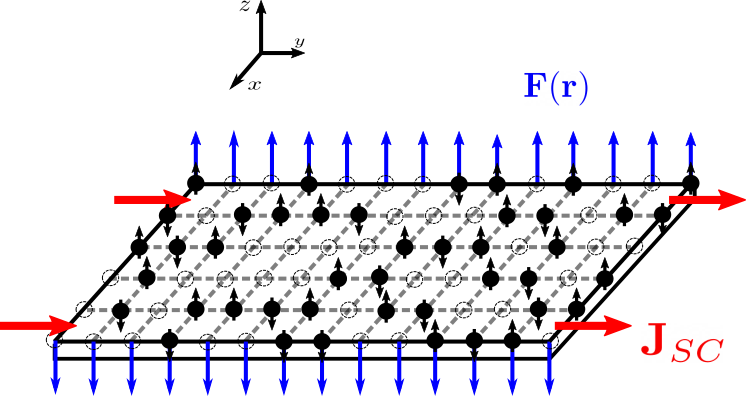}
	\caption{Origin of flux noise. Section of a superconducting wire with spin impurities randomly distributed at its surface. The flux produced by each spin is given by $\hat{\Phi}_i=-\bm{F}(\bm{R}_i)\cdot \bm{\hat{s}}_i$, where $\bm{\hat{s}}_i$ is the spin operator of an impurity located at $\bm{r}=\bm{R}_i$. The ``flux vector" $\bm{F}(\bm{r})$ points along the magnetic field produced by the current density $\bm{J}_{SC}$ shown in the figure. Spin impurity dynamics leads to background flux noise that limits coherence times for SQUID-based and flux-tunable qubits.} 
\label{FigFlux}	
\end{figure}

The origin of flux noise in superconducting devices remains unknown, although there is consensus that it arises from the dynamics of spin centers (magnetic impurities) near the superconducting wires \cite{deSousa2007,Koch2007a, Sendelbach2008, Faoro2008, Kumar2016, deGraaf2017, Quintana2017} (see Fig.~\ref{FigFlux}).
This conclusion is supported by experiments showing noise amplitude following the Curie susceptibility law ($\tilde{\chi}(\omega=0)\propto 1/T$, where $T$ is temperature)  \cite{Sendelbach2008}. However, there are conflicting opinions about the magnetic order of the spins causing noise. While some authors claimed proximity to a spin glass phase \cite{Sendelbach2008, Lanting2014a}, recent experiments were able to measure a Curie-Weiss susceptibility $\tilde{\chi}(\omega=0)\propto 1/(T-T_{{\rm CW}})$ which rules out the spin glass scenario. Instead, \cite{Quintana2017} measured $T_{{\rm CW}}\approx -10~{\rm mK}<0$ indicating proximity  to an \emph{antiferromagnetic} phase, while \cite{Lanting2020} measured $T_{CW}\approx +5~{\rm mK}>0$ indicating proximity to a \emph{ferromagnetic} phase. 

Measurements of flux noise \cite{Bylander2011, Anton2013, Lanting2014a, Quintana2017} are quite puzzling. They reveal approximate $\tilde{S}_{\Phi}(\omega)\propto 1/\omega^{\alpha}$ frequency dependence over several decades of frequency, %($\omega=2\pi \times 10^{-4}-10^{9}$~Hz), 
and show quantum-noise asymmetry $\tilde{S}_{\Phi}^{-}(\omega)=\tilde{S}_{\Phi}(\omega)-\tilde{S}_{\Phi}(-\omega)$ due to spontaneous emission \cite{Quintana2017}. To our knowledge all theories of flux noise available in the literature are semiclassical so they are not able to describe this asymmetry \cite{Faoro2008, Lanting2014a, Atalaya2014}. Moreover, flux noise was shown to become either Ohmic $\propto \omega$ \cite{Lanting2011} or super-Ohmic $\propto \omega^{3}$ above $4$~GHz \cite{Yan2016, Quintana2017}, and a key open question is whether or not this high frequency contribution is also due to impurity spins. 

The main mechanism for spin energy relaxation in ferromagnetic metals, the so called Gilbert damping, relies on magnetic excitations (magnons) decaying into electron-hole pair excitations in the metal. In a superconductor these are exponentially suppressed at temperatures much lower than the superconducting energy gap, making the Gilbert damping constant exponentially small \cite{Bell2008}. In addition, the weak magnetic fields in superconducting devices implies the spin-orbit and hyperfine spin-flip rates are close to zero \cite{deSousa2007}.
The only remaining mechanism for spin energy decay in superconductors is due to the interaction between each impurity spin with nearby amorphous two-level systems \cite{deSousa2007}. Such an interaction leads to wide distributions of single-spin-flip rates $\Gamma_i$ for different impurity spins $i$ \cite{Belli2020}. 

It is known in nuclear magnetic resonance experiments that in the presence of spin-spin interaction, spins with $\Gamma_i=0$ can relax by diffusing their nonequilibrium magnetization towards sites with $\Gamma_i >0$ \cite{Bloembergen1949}.  However, no theory to date has been able to capture the coexistence of spin-spin interaction with wide distributions of $\Gamma_i$. 

When $\Gamma_i=0$ for all spins $i$, the total spin magnetization is conserved, and the spin fluctuations due to spin-spin exchange interaction necessarily obey a spin diffusion equation at long wavelengths \cite{DeGennes1958, Bennett1965}. 
A recent ``pump and probe" experiment \cite{Lanting2020} measured the flux time correlation function $\langle \hat{\Phi}(t)\hat{\Phi}(0)\rangle$ in SQUIDs, and showed that it behaved similar to Brownian motion: $\langle \Phi(t)\Phi(0)\rangle/\langle\Phi^2\rangle \propto 1-{\rm const.}\sqrt{t}$ in the $1-1000$~$\mu$s time range. It provides evidence that the dynamics of flux $\Phi$ in a superconducting device is described by the phenomena of diffusion in this time range. 

The usual theory for spin dynamics in a disordered spin system is based on the assumption that the system is in a spin glass phase, that is uniform and translational invariant \cite{Hertz1983}. 
These theories are not satisfactory for modeling flux noise in superconducting devices for two reasons. First, there is evidence that the impurity spins are in the paramagnetic (non-spin glass) phase \cite{Quintana2017, Lanting2020}. Second, inhomogeneity and lack of translation invariance play a crucial role. For example, the flux produced by spins located close to the superconducting wire edge is much larger than the flux produced by spins away from the edges  \cite{Koch2007a, LaForest2015}. There is also the desire to know what is the impact of nonhomogeneous spin distributions, such as impurity spin clustering \cite{Anton2013, Atalaya2014}.

Describing spin diffusion from ``first principles'', i.e. by integrating the Heisenberg equations of motion for a model of interacting spins, is a well known challenge of theoretical physics \cite{Bennett1965}. 
The standard method is what we call ``third-principles approach'': It assumes the spins can be described by a continuous magnetization density $\bm{M}(\bm{r},t)$ that satisfies the phenomenological equation $\frac{\partial \bm{M}}{\partial t}=D\nabla^{2}\bm{M}$, with $D$ the spin diffusion constant. With all physical properties lumped into a phenomenological constant $D$, the third-principles theory can not establish a connection to microscopic model spin Hamiltonians, spin spatial distributions, and the impact of inhomogeneity and clusters. 

A serious shortcoming of the third principles approach is that it requires the assumption of a hard boundary condition  such as $\bm{M}(\bm{r},t)=\bm{0}$ at wire edges and the surface of spin clusters in order to ensure total spin conservation across the boundaries \cite{Faoro2008, Lanting2020}. A hard boundary condition like this is unjustifiable, and in fact is known to be violated in magnetic systems due to the appearence of confined surface or edge magnons \cite{Beairsto2021}. Developing a theory of spin dynamics that properly accounts for the boundary effects is of crucial importance because flux noise is known to be dominated by spins at the edge of the wire where the supercurrent is maximum \cite{Koch2007a, LaForest2015}.

The purpose of this article is to propose a ``second principles'' theoretical framework for spin dissipation (diffusion plus relaxation) that includes quantum noise and is more ``microscopic'' than the usual third-principles approach. The goal is to establish a connection between flux noise measured in experiments and microscopic spin Hamiltonians, without the prohibitive computational cost associated to the first-principles approach. \Ch{To do this we assume spin dissipation according to a random walk model governed by the
parameters of the spin Hamiltonian such as the microscopic exchange interaction between each pair of spins}. 

Below we describe general theoretical results, and then present explicit numerical calculations of flux noise for the Heisenberg model with nearest neighbour interactions in the paramagnetic phase ($T>T_{c}^{{\rm mag}}$). 
Our calculations are done in a finite spin lattice with a random distribution of vacancies, showing explicit predictions for spatial confinement (wire edges) and disorder due to random distribution of vacancies across the wire's surface as well as wide distributions of individual spin-flip rates $\Gamma_i$.

\section{Model for flux noise and linear response theory \label{sec:model}}

We start by describing the impact of wire currents on impurity spins and how it leads to a general expression for flux noise.  The magnetic moment of an impurity spin is given by $-g\mu_B \hat{\bm{s}}_i$, where $g\approx 2$ is the $g$-factor, $\mu_B$ is the Bohr magneton, and $\hat{\bm{s}}_i$ is a dimensionless spin operator for an impurity located at position $\bm{R}_i$. It couples to the superconducting wire current density by producing a flux \cite{LaForest2015},
\beq
\hat{\Phi}=-\sum_i x_i\bm{F}(\bm{R}_i)\cdot \hat{\bm{s}}_i, 
\eeq
where the sum goes over all sites $\bm{R}_i$ of a virtual square lattice containing $N$ sites. The variable $x_i=1$ when there is a spin at the virtual site, and $x_i=0$ otherwise, and the spin density is $\sigma=\sum_i x_i/N=N_s/N$, where 
$N_s$ is the number of spins. 
The flux vector $\bm{F}(\bm{r})$ is directly proportional to $\bm{B}_{I}(\bm{r})$, the magnetic field produced by the wire's current density:
$\bm{F}(\bm{r})=g\mu_B\bm{B}_{I}(\bm{r})/I$, where $I$ is the total current flowing through the wire. 

The wire's current in turn affect spins by imprinting an \emph{external} local field 
\begin{equation} \label{eq_hfield}
\bm{h}_i = - g \mu_B \bm{B}_{I} (\bm{R}_i) = - I \bm{F}(\bm{R}_i),
\end{equation}
that couples to the spin according to ${\cal H}_c = - \sum_i x_i \bm{h}_i \cdot \hat{\bm{s}}_i$, so $\bm{h}_i$ has dimensions of energy. 
When the local field $\bm{h}_i$ is time dependent, the spins respond according to
\begin{equation} \label{eq_lr_dynamic}
\langle \hat{s}_i^{a}(t) \rangle_{h \neq 0} = \langle \hat{s}_i^{a}(t) \rangle_{h = 0} + \sum_{j,b} \int_{- \infty}^{\infty} dt' \chi_{ij}^{a b}(t-t') h_j^{b}(t'),
\end{equation}
where $a,b=x,y,z$ and the dynamical susceptibility is given by the linear response formula,
\begin{equation} \label{eq_chi}
\chi^{a b}_{ij} (t-t') = x_i x_j \frac{i}{\hbar} \theta (t-t') \langle [\hat{s}_i^{a}(t), \hat{s}_j^{b}(t')]\rangle.
\end{equation}
Defining spin noise as
\begin{equation}
\tilde{S}^{ab}_{ij} (\omega) = x_i x_j \int_{-\infty}^{\infty} dt e^{i \omega t} \langle [\hat{s}_i^{a}(t) - \langle \hat{s}_i^{a} \rangle] [\hat{s}_j^{b}(0) - \langle \hat{s}_j^{b} \rangle]\rangle,
\end{equation}
and using Eq. (\ref{eq_chi}) we obtain the general relationship between susceptibility and spin noise,
\begin{equation}
\label{suscnoise}
\tilde{\chi}^{ab}_{ij}(\omega) = \frac{1}{2 \pi \hbar} \int_{-\infty}^{\infty} d\omega' \frac{1 - e^{- \frac{\hbar \omega'}{ k_B T}}}{\omega' - \omega - i \eta} \tilde{S}^{ab}_{ij} (\omega'),
\end{equation}
where $\eta \rightarrow 0^+$. Taking the imaginary part and using the fact that $\tilde{S}^{ab}_{ij} (\omega) = \tilde{S}^{ba}_{ji} (\omega)^{*}$, we get the fluctuation-dissipation theorem for spins,
\begin{equation} \label{eq_FDT}
\tilde{S}^{ab}_{ij} (\omega) = \frac{2 \hbar}{1 - e^{-\hbar \omega / k_B T}} \frac{1}{2i} [\tilde{\chi}^{ab}_{ij} (\omega) - \tilde{\chi}^{ba}_{ji} (\omega)^{*}].
\end{equation}
The flux noise is then given by
\begin{eqnarray} \label{eq_fluxnoise}
\tilde{S}_{\Phi} (\omega) &=& \int_{-\infty}^{\infty}dt \textrm{e}^{i\omega t} \left\langle \delta \hat{\Phi}(t)\delta\hat{\Phi}(0)\right\rangle\nonumber\\
&=&\sum_{i,j,a,b} F^{a}(\bm{R}_i) \tilde{S}^{ab}_{ij} (\omega) F^{b}(\bm{R}_j),
\end{eqnarray}
where $\delta\hat{\Phi}(t)=\hat{\Phi}(t)-\langle \hat{\Phi}\rangle$ denotes flux fluctuation. 

\section{Theory of spin dynamics\label{section:model_spins}}

Our goal is to compute flux noise for the general Heisenberg quantum spin Hamiltonian,
\beq
{\cal H}=-\frac{1}{2}\sum_{i,j}x_i x_j J_{ij}\hat{\bm{s}}_i\cdot\hat{\bm{s}}_j - \sum_i x_i \bm{h}_i \cdot \hat{\bm{s}}_i,
\label{spin_hamiltonian}
\eeq
plus spin energy relaxation due to other degrees of freedom such as phonons, electron-hole excitations, and two-level system defects \cite{Belli2020}. Here $J_{ij}$ is the exchange interaction between spins $i$ and $j$, that can be ferromagnetic ($J_{ij}>0$) or antiferromagnetic ($J_{ij}<0$), and $\bm{h}_i$ is the local field defined in Eq.~(\ref{eq_hfield}).

\subsection{Static mean-field theory}

In mean-field theory we neglect higher order fluctuations by approximating $\langle \hat{s}_{i}^{a} \hat{s}_{j}^{b} \rangle \approx \langle \hat{s}_{i}^{a} \rangle \langle \hat{s}_{j}^{b} \rangle$. We simplify the notation by writing $\langle \hat{\bm{s}}_i \rangle = \bm{s}_i$, i.e. the spin vector without a hat denotes the average of the spin operator (a real vector).
The mean-field approximation is exactly the same as the ``classical spin model'' used by many authors, e.g. \cite{Atalaya2014}. An additional approximation in mean-field theory is to assume the system's entropy can be written as a sum of single-spin entropies \cite{Chaikin1995}:
\begin{equation}
\langle {\cal S} \rangle = k_B \sum_i x_i \left[ \ln{2} - 2|\bm{s}_i|^2 - \frac{4}{3} (|\bm{s}_i|^2)^2 + \mathcal{O}((|\bm{s}_i|^2)^3) \right].
\end{equation}
This expression is specific to spin-1/2 impurities. Note that cutting the expansion to fourth order affects the result only when $T \ll T_{c}^{{\rm mag}}$, where $T_{c}^{{\rm mag}}$ is a critical temperature for a phase where $s_i > 0$. The free energy is thus given by:
\begin{eqnarray}
{\cal F} &=& \langle {\cal H} \rangle - T \langle {\cal S} \rangle = -\frac{1}{2} \sum_{i,j} x_i x_j J_{ij} \bm{s}_i \cdot \bm{s}_j - \sum_i x_i \bm{h}_i \cdot \bm{s}_i \nonumber\\ 
&&- (k_B T) \sum_i x_i \left[ \ln{2} - 2|\bm{s}_i|^2 - \frac{4}{3} (|\bm{s}_i|^2)^2\right]. 
\label{free_energy}
\end{eqnarray}
Usually thermal equilibrium is realized by the set of $\bm{s}_i$ that leads to the global minimum of the free energy. For example, take $x_i=1$ for all $i$, $J_{ij}=J>0$ for nearest neighbors and zero otherwise, and a lattice with periodic boundary conditions (b.c.). 
In this case the global minimum of ${\cal F}$ is realized by $\bm{s}_i=\bm{s}^{{\rm eq}}$ for all $i$ (the ferromagnetic homogeneous state), with free energy given by
\beq
\frac{{\cal F}}{N}= \left(2k_BT -\frac{z J}{2}\right)(s^{{\rm eq}})^{2}+\frac{4}{3}k_BT (s^{{\rm eq}})^{4}-k_BT\ln{2},
\eeq 
where $z$ is the number of nearest neighbours for each site of the lattice. From this expression we see that a global minimum with $s^{{\rm eq}}>0$ appears only when the first term changes sign, leading to critical temperature $k_B T_{c}^{{\rm mag}}=k_BT_{{\rm CW}}=z J/4$. 
The same calculation can be done for $J<0$ when the lattice can be partitioned into two sublattices with one being n.n. to the other. For this case the global minimum of ${\cal F}$ is realized by $\bm{s}_i=+\bm{s}^{{\rm eq}}$ for one sublattice and $\bm{s}_i=-\bm{s}^{{\rm eq}}$ for the other (antiferromagnetic homogeneous state). This leads to $k_BT_c=z |J|/4$. However, $k_BT_{{\rm CW}}=-z |J|/4$ because the magnetic susceptibility does not have a singularity at $T_c$ (it's the staggered susceptibility that is singular at $T_c$). These are the well-known mean-field results for phase transitions in the spin-$1/2$ Heisenberg model \cite{Chaikin1995}. 

\subsection{Dynamical mean-field theory \label{section:dynamical_mean_field}}

Based on fundamental theories of spin dynamics \cite{Chaikin1995} we propose the following generalized equation of motion for the spins, 
\begin{equation} \label{eom}
    \frac{d \bm{s}_i}{d t} =  \frac{1}{\hbar} \bm{s}_i \times \bm{H}_i - \sum_{j}D_{ij} \bm{H}_j - \Gamma_{i}(\bm{s}_i - \bm{s}_{i}^{{\rm inst~eq}}),
\end{equation}
expected to be valid for frequencies smaller than a cut-off $\Omega_c$ to be discussed later. 
In addition to the usual spin precession, this includes a discrete version of the \emph{intra-spin dissipation} operator $D_{ij}$,
together with isotropic \emph{spin energy relaxation} $\Gamma_{i}$. 
The $\Gamma_i$ drives $\bm{s}_i$ towards its ``instantaneous equilibrium'' value 
\begin{equation}
    \bm{s}_{i}^{{\rm inst~eq}}=\bm{s}_{i}^{{\rm eq}}+\sum_j\tilde{\chi}_{ij}(0)\delta\bm{h}_j(t),
\label{instapprox}
\end{equation}
which is time-dependent due to local field dynamics, $\bm{h}_i(t)=\bm{h}_{i}^{{\rm eq}}+\delta\bm{h}_i(t)$, where $\bm{h}_{i}^{{\rm eq}}$ is the static part. The quantities $\bm{s}_{i}^{{\rm eq}}$ do not depend on time, they are thermal equilibrium averages calculated assuming $\delta\bm{h}_i(t)=0$, i.e. they only depend on $\bm{h}_{i}^{{\rm eq}}$ and other static free energy parameters;
$\tilde{\chi}_{ij}^{ab}(0)=\partial (s_{i}^{{\rm eq}})^{a}/\partial h_{j}^{b}$ is the $\omega=0$ susceptibility, assumed isotropic ($\propto \delta_{ab}$) to be consistent with our Hamiltonian (\ref{spin_hamiltonian}).
We call Eq.~(\ref{instapprox}) the ``instantaneous approximation'', because it assumes the other degrees of freedom causing spin energy relaxation relax much faster than the spins themselves, so that the spin system remains in thermal equilibrium with the other non-spin degrees of freedom at all times. 
Note how Eq.~(\ref{instapprox}) introduces the $\omega=0$ susceptibility \emph{self-consistently} into the equation of motion~(\ref{eom}).

%The equilibrium spin magnetization $\bm{s}_{i}^{{\rm eq}}$ in Eq.~(\ref{eom}) is assumed to remain in thermal equilibrium with the other non-spin degrees of freedom at all times, so that $\bm{s}_{i}^{{\rm eq}}\equiv\bm{s}_{i}^{{\rm eq}}(\bm{h}_{i}(t))$. We call this the ``instantaneous approximation'', because it assumes the other degrees of freedom causing spin energy relaxation relax much faster than the spin itself. This approximation makes $\bm{s}_{i}^{{\rm eq}}$ time-dependent according to $\bm{s}_{i}^{{\rm eq}}=\bm{s}_{i}^{{\rm eq}}(\bm{h}_{i}^{{\rm eq}})+\sum_j\tilde{\chi}_{ij}(0)\delta\bm{h}_j(t)$, where $\tilde{\chi}_{ij}^{\alpha \beta}(0)=\partial s_{i}^{eq \alpha}/\partial h_{j}^{\beta}$ is the $\omega=0$ susceptibility, assumed isotropic ($\propto \delta_{\alpha\beta}$) to be consistent with our Hamiltonian (\ref{spin_hamiltonian}). From now on we denote $\bm{s}_{i}^{{\rm eq}}(\bm{h}_{i}^{{\rm eq}})\equiv \bm{s}_{i}^{{\rm eq}}$ below.

%Since $J_{ij}=J_{ji}$, we also assume $D_{ij}=D_{ji}$ (i.e., the matrix $\Mat{D}$ is symmetric). 
The \emph{internal} spin field is defined as
\begin{eqnarray} 
\label{internal_spin_field}
    \bm{H}_i &=& - \frac{\partial \mathcal{F}}{\partial \bm{s}_i}\\
&=& x_i \left\{ \sum_j x_j J_{ij} \bm{s}_j
+\bm{h}_i-4k_BT\left[1+\frac{4}{3}s_{i}^{2}\right]\bm{s}_i \right\},\nonumber
\end{eqnarray}
and the thermal equilibrium spin averages $\bm{s}_{i}^{{\rm eq}}$ are determined by imposing time independence, $\delta \bm{h}_i(t)=\bm{0}$ and 
$\frac{d \bm{s}_{i}}{d t}=\bm{0}$ for all $i$. This implies $\bm{s}_{i}^{{\rm eq}}$ must be found by solving 
the system of equations
\beq
\frac{1}{\hbar}\bm{s}_{i}^{{\rm eq}}\times \bm{H}_{i}^{{\rm eq}}-\sum_{j}D_{ij} \bm{H}_j^{{\rm eq}}=\bm{0}, 
\label{eq_equations}
\eeq
where $\bm{H}_{i}^{{\rm eq}}$ is Eq.~(\ref{internal_spin_field}) with $\bm{s}_i=\bm{s}_{i}^{{\rm eq}}$ and $\bm{h}_i=\bm{h}_{i}^{{\rm eq}}$.  Note how Eq.~(\ref{eq_equations}) is always satisfied for $\bm{H}_{i}^{{\rm eq}}=\bm{0}$, a smooth local minimum of the free energy. However, other solutions with $\bm{H}_{i}^{{\rm eq}}\neq\bm{0}$ may arise in the presence of site-dependent local fields $\bm{h}_{i}^{{\rm eq}}$.

The three terms in the right hand side of Eq.~(\ref{eom}) correspond to reactive dynamics, intra-spin-system dissipation (e.g. diffusion), and spin energy relaxation due to other degrees of freedom, respectively. The reactive term is non-dissipative, it does not change sign under time reversal $t \rightarrow -t$ so it has the same symmetry as the left hand side. The second and third terms on the right hand side do change sign under time reversal, leading to an irreversible approach to thermal equilibrium (the arrow of time). These terms must be added \textit{ad hoc} to the linearized equations of motion so that the zeroth (attainment of thermal equilibrium) and second (entropy always increases) laws of thermodynamics are obeyed. That is, the system is able to reach thermal equilibrium, and the free energy always decreases as a function of time when the system is in contact with a thermal reservoir. 

A few notes about the microscopic origin of spin dissipation $D_{ij}$ are warranted. The normal modes of Eq.~(\ref{eom}) 
are called \emph{magnons} and \emph{paramagnons}, to be defined below. The reactive terms of Eq.~(\ref{eom}) describe the dynamics of noninteracting (para)magnons obtained by the mean-field approximation. Exactly the same results are obtained from different methods, e.g. using a Holstein-Primakoff transformation to convert spin operators into Bosonic creation/destruction operators; transforming Hamiltonian (\ref{spin_hamiltonian}) and keeping contributions that are quadratic in these Bosonic operators leads to the same magnon modes as Eq.~(\ref{eom}) with $D_{ij}=\Gamma_i=0$ \cite{Beairsto2021}. However, the higher order terms that are neglected in this quadratic approximation can be interpreted as describing (para)magnon-(para)magnon interactions. The introduction of $D_{ij}\neq 0$ serves to account for these interactions phenomenologically. 

\section{Specification of dissipation matrix 
\texorpdfstring{$D_{ij}$}{} in the presence of confinement and disorder}

To go beyond the third-principles assumption of long wavelength spin diffusion we need to come up with a specification for $D_{ij}$ that respects several physical constraints. To do this, we take inspiration from random walk models in a lattice. The key idea is that exchange interaction $J_{ij}$ is the main driver for each random walk step, a spin ``flip-flop''. A sequence of many flip-flops will lead to diffusion. The constraint of total spin conservation motivates our postulation of  the following spin dissipation matrix:
\begin{equation} \label{diff_operatorDij}
    D_{ij} = \frac{d_0(T)}{\hbar\bar{J}_c} \left( x_i x_j |J_{ij}| -  \delta_{ij} \sum_k x_i x_k |J_{ik}|  \right).
\end{equation}
Here $d_0(T)$ is a function of temperature to be determined by fitting the theory to experiments (note $d_0(T)$ is dimensionless). This is introduced to account for critical behaviour of the spin diffusion constant near $T_{c}^{{\rm mag}}$ \cite{Halperin1967}.
The quantity
\beq
\bar{J}_c = \frac{1}{N_{c}} \sum_{j,k\in {\rm cluster~}c}x_j x_k |J_{jk}|
\label{eq_jc}
\eeq
is the average exchange times coordination number for $c^{th}$ cluster, the cluster that contains spin $i$. Such a cluster is defined as the set of all spins $j$ such that either $J_{ij}\neq 0$ or there exists a set of sites $k_1, k_2, \ldots, k_n$ such that $J_{ik_1}J_{k_1k_2}J_{k_2k_3}\cdots J_{k_{n}j}\neq 0$. $N_c$ is the number of spins in the isolated cluster $c$.  
The following motivates this choice:
\begin{enumerate}
\item When the local external field $\bm{h}_i$ and spin relaxation rate $\Gamma_{i}$ are both zero, Eq.~(\ref{diff_operatorDij}) preserves total spin. For each isolated spin cluster, $\sum_{i,j\in {\rm cluster}}D_{ij}\bm{H}_{j}=\bm{0}$,  therefore summing Eq.~(\ref{eom}) over all spins in a cluster leads to $\frac{d}{d t} (\sum_{i\in {\rm cluster}} \bm{s}_i) = \bm{0}$, so the total spin in each cluster is a constant of the motion. 
Therefore, we do not need to assume hard boundary conditions \cite{Faoro2008, Lanting2020} to describe confined systems such as spin clusters and wire edges.  

\item This choice for $D_{ij}$ gives rise to diffusion in the long wavelength regime. 
E.g. for the homogeneous nearest-neighbor model in the square lattice with 
$J_{ij} = J_{i, i + \bm{v}} = J$ for $\bm{v} = \pm a \hat{x}, \pm a \hat{y}$, we get
\begin{equation}
-\sum_{i,j} \delta(\bm{r}-\bm{R}_i)D_{ij} \bm{H}_j= 
\frac{d_0(T) a^2 k_BT}{\hbar}\nabla^2 \bm{M}(\bm{r}),
\label{cont_limit}
\end{equation}
where we assumed high temperature ($k_B T \gg J$ and $(\bm{s}_i)^2 \ll 1$) and took the continuum limit by defining the magnetization density $\bm{M}(\bm{r}) = - \sum_i \bm{s}_i \delta (\bm{r} - \bm{R}_i)$. 
Equation~(\ref{cont_limit}) may be compared to experiments that show spin diffusion constant increasing with temperature \cite{Lanting2014a}. 

\item Consider the time derivative of the free energy
%$\delta \bm{h}_i(t)=\bm{0}$ 
in each isolated spin cluster when $\bm{h}_i$ is independent of time:
\begin{widetext}
\begin{eqnarray} \label{eq_free_energy_timederivative}
        \frac{d \mathcal{F}}{d t} &=&\sum_i \frac{\partial \mathcal{F}}{\partial \bm{s}_i} \cdot \frac{d \bm{s}_i}{d t}
        = \frac{d_0(T)}{\hbar\bar{J}_c}  \sum_i \bm{H}_i \cdot \sum_j x_i x_j |J_{ij}| (\bm{H}_j - \bm{H}_i) + \sum_i \Gamma_i \bm{H}_i \cdot (\bm{s}_i - \bm{s}_i^{{\rm eq}})\nonumber\\
        &=& - \frac{d_0(T)}{\hbar\bar{J}_c} \sum_{i<j} x_i x_j |J_{ij}| (\bm{H}_i - \bm{H}_j)^2 - \sum_i \Gamma_i \left( \mathcal{F}(\bm{s}_i) - \mathcal{F}(\bm{s}_i^{{\rm eq}}) + \mathcal{O} \left[ (\delta \bm{s}_i)^2 \right] \right).
\end{eqnarray}
\end{widetext}
The first term on the RHS is always negative, showing that our choice for $D_{ij}$ always tends to decrease the free energy as time increases (i.e. it obeys the 2nd law of thermodynamics). This justifies our use of modulus of $J_{ij}$ in Eq. (\ref{diff_operatorDij}). 

The second term on the RHS of Eq. (\ref{eq_free_energy_timederivative}) is negative provided that the deviation out of equilibrium is small and $\mathcal{F}(\bm{s}_i^{{\rm eq}})$ is a local minimum of the free energy. Therefore, the coupling to other nonspin degrees of freedom $\Gamma_i$ pushes the system towards a local  minimum of the free energy, without subjecting to spin conservation. 

\end{enumerate}

\section{Calculation of dynamical susceptibility}

For small deviations from equilibrium we write 
$\bm{h}_i=\bm{h}_{i}^{{\rm eq}}+\delta\bm{h}_i(t)$, and $\bm{s}_i = \bm{s}_i^{{\rm eq}} + \delta \bm{s}_i(t)$, where both $\delta\bm{h}_{i}(t)$ and $\delta\bm{s}_i(t)$ are small time-dependent perturbations.
%; $\bm{h}_{i}^{{\rm eq}}$ is the time-independent external field due to superconducting (SC) currents, which is not necessarily small. 

%The equilibrium spin magnetization $\bm{s}_{i}^{{\rm eq}}$ in Eq.~(\ref{eom}) is assumed to remain in thermal equilibrium with the other non-spin degrees of freedom at all times, so that $\bm{s}_{i}^{{\rm eq}}\equiv\bm{s}_{i}^{{\rm eq}}(\bm{h}_{i}(t))$. We call this the ``instantaneous approximation'', because it assumes the other degrees of freedom causing spin energy relaxation relax much faster than the spin itself. This approximation makes $\bm{s}_{i}^{{\rm eq}}$ time-dependent according to $\bm{s}_{i}^{{\rm eq}}=\bm{s}_{i}^{{\rm eq}}(\bm{h}_{i}^{{\rm eq}})+\sum_j\tilde{\chi}_{ij}(0)\delta\bm{h}_j(t)$, where $\tilde{\chi}_{ij}^{\alpha \beta}(0)=\partial s_{i}^{eq \alpha}/\partial h_{j}^{\beta}$ is the $\omega=0$ susceptibility, assumed isotropic ($\propto \delta_{\alpha\beta}$) to be consistent with our Hamiltonian (\ref{spin_hamiltonian}). From now on we denote $\bm{s}_{i}^{{\rm eq}}(\bm{h}_{i}^{{\rm eq}})\equiv \bm{s}_{i}^{{\rm eq}}$ below.

We plug these into the equation of motion (\ref{eom}), drop non-linear terms such as $(\delta \bm{s}_i)^2, (\delta\bm{h}_{i}(t))^{2}$, and use Eq.~(\ref{eq_equations}) to simplify:
\begin{widetext}
\begin{eqnarray}
\label{eq_eom_linear}
    \frac{d}{d t} (\delta \bm{s}_i) &=& \frac{1}{\hbar} \bm{s}_i^{{\rm eq}} \times \left\{(\delta\bm{h}_i)+\sum_j x_j J_{ij} (\delta \bm{s}_j)-4 k_B T \left[1 + \frac{4}{3}(s_{i}^{{\rm eq}})^2\right](\delta \bm{s}_i) - \frac{32}{3} k_B T [\bm{s}_{i}^{{\rm eq}} \cdot (\delta \bm{s}_i)] \bm{s}_i^{{\rm eq}}\right\} \nonumber\\
    &&-\sum_j D_{ij} x_j \left\{(\delta\bm{h}_j)+\sum_k x_k J_{jk} (\delta \bm{s}_k)-4 k_B T \left[1 + \frac{4}{3}(s_{j}^{{\rm eq}})^2\right](\delta \bm{s}_j) - \frac{32}{3} k_B T [\bm{s}_{j}^{{\rm eq}} \cdot (\delta \bm{s}_j)] \bm{s}_j^{{\rm eq}}\right\}\nonumber\\
    &&+\frac{1}{\hbar}(\delta \bm{s}_i)\times \bm{H}_{i}^{{\rm eq}}-\Gamma_i (\delta \bm{s}_i)+\Gamma_i\sum_j\tilde{\chi}_{ij}(0)(\delta \bm{h}_{j}).
\end{eqnarray}
\end{widetext}
We assume $\bm{s}_i^{{\rm eq}} = s_i^{{\rm eq}} \hat{\bm{z}}$ and $\bm{H}^{{\rm eq}}_{i}=H^{{\rm eq}}_{i}\bm{\hat{z}}$ and break this down into two equations, one for $\delta s_i^z$ obtained by dot product with $\bm{\hat{z}}$ on both sides of Eq.~(\ref{eq_eom_linear}), 
and the other for $\delta s_i^{+} =\delta s^{x}_i+i\delta s^{y}_i$ obtained by dot product with $(\bm{\hat{x}}+i\bm{\hat{y}})$. 
Taking the time Fourier transform we get two decoupled equations:
\begin{subequations}
\begin{eqnarray}
    \label{zEOM}
    (\omega \bm{{\sf I}} - \bm{{\sf P}}) \cdot \delta \tilde{\bm{s}}^z &=&  i \left(\Mat{\Gamma}\cdot\Mat{\tilde{\chi}}_0 -\Mat{D}\right) \cdot\delta\tilde{\bm{h}}^z,\\
    \label{plusEOM}
    (\omega \bm{{\sf I}} - \bm{{\sf M}}) \cdot \delta \tilde{\bm{s}}^+ &=&  \left[i \left(\Mat{\Gamma}\cdot\Mat{\tilde{\chi}}_0 -\Mat{D}\right)  -\frac{1}{\hbar}\bm{{\sf s}}^{{\rm eq}}\right]\cdot \delta\tilde{\bm{h}}^{+},
\end{eqnarray}
\end{subequations}
where $\delta\tilde{\bm{s}}^{z}, \delta\tilde{\bm{s}}^{+}$ and $\delta\tilde{\bm{h}}^{z}, \delta\tilde{\bm{h}}^{+}$ are $N$-component column vectors, and $\Mat{I}, \Mat{\Gamma}, \Mat{\tilde{\chi}}_0,\Mat{D}, \Mat{s}^{{\rm eq}}$ are $N\times N$ matrices.  They are defined by $[\Mat{I}]_{ij}=x_i \delta_{ij}$, $[\Mat{\Gamma}]_{ij}=x_i\Gamma_i\delta_{ij}$, 
$[\Mat{\tilde{\chi}}_0]_{ij}=\tilde{\chi}_{ij}(0)$, $[\Mat{D}]_{ij}=x_ix_j D_{ij}$, and $[\Mat{s}^{{\rm eq}}]_{ij}=x_is^{{\rm eq}}_{i}\delta_{ij}$.

The matrices $\Mat{P}$ and $\Mat{M}$ are the \emph{paramagnon} and \emph{magnon} matrices, respectively. They are given by 
\begin{subequations}
    \begin{eqnarray}
        \label{mat_P}
        \Mat{P}&=&-i \left\{\Mat{\Gamma}+\Mat{D}\cdot\Mat{J}-4k_BT \Mat{D}\cdot\left[\Mat{I}+4\left(\Mat{s}^{{\rm eq}}\right)^{2}\right]\right\},\\
        \label{mat_M}
        \Mat{M}&=&-i \left\{\Mat{\Gamma}+\Mat{D}\cdot\Mat{J}-4k_BT \Mat{D}\cdot\left[\Mat{I}+\frac{4}{3}\left(\Mat{s}^{{\rm eq}}\right)^{2}\right]
        \right\}\nonumber\\
        +&\frac{1}{\hbar}&\left\{\Mat{H}^{{\rm eq}}-\Mat{s}^{{\rm eq}}\cdot \left[\Mat{J}-4k_BT\left(\Mat{I}+\frac{4}{3}\left(\Mat{s}^{{\rm eq}}\right)^{2}\right)\right]\right\},
    \end{eqnarray}
\end{subequations}
where $[\Mat{J}]_{ij}=x_ix_j J_{ij}$, and 
$[\Mat{H}^{{\rm eq}}]_{ij}=H^{{\rm eq}}_{i}\delta_{ij}$.

The eigenvalues of $\Mat{P}$ and $\Mat{M}$ are paramagnon and magnon frequencies, respectively. While paramagnons cause spin fluctuations along $\bm{s}_{i}^{{\rm eq}}$, the magnons cause fluctuations perpendicular to $\bm{s}_{i}^{{\rm eq}}$. In the paramagnetic phase ($T>T_{c}^{{\rm mag}}$) with zero external fields, $\Mat{s}^{{\rm eq}} = \Mat{H}^{{\rm eq}}= \Mat{0}$ and the matrices $\Mat{P}$ and $\Mat{M}$ become identical, signaling the presence of isotropic spin fluctuations (i.e. paramagnons are three-fold degenerate). From now on we shall focus our discussion on this paramagnetic regime, so we only need to consider the spectrum of $\Mat{P}$. 

Diagonalize $\bm{{\sf P}}$ with a transformation $\bm{{\sf U}}$ such that
\beq
    \label{trans_P}
    \bm{{\sf U}}^{-1} \cdot\bm{{\sf P}}\cdot \bm{{\sf U}} = \bm{{\sf P}}_d = -i\sum_{m} \gamma_{m} \bm{\hat{e}}_{m} \otimes \bm{\hat{e}}_{m}^{T},
\eeq
where $m$ labels the paramagnon mode with frequency $-i\gamma_m$, with $\bm{\hat{e}}_m$ unit column vectors, $\bm{\hat{e}}_{m}^{T}=(0,\ldots,0,1,0,\ldots,0)$, etc. 
All elements of $\Mat{P}$ are pure complex, therefore the $\Mat{U}$ and $\Mat{U}^{-1}$ can be chosen to have real elements. Take the complex conjugate of Eq.~(\ref{trans_P}) and use $\Mat{P}^{*}=-\Mat{P}$ to see that $\gamma^{*}_{m}=\gamma_{m}$.

Apply $\Mat{U}^{-1}$ on both sides of Eq.~(\ref{zEOM}),
\beq
\left(\omega-\Mat{P}_d\right)\cdot\Mat{U}^{-1}\cdot \delta\tilde{\bm{s}}^{z}=i\Mat{U}^{-1}\cdot \left(\Mat{\Gamma}\cdot\Mat{\tilde{\chi}}_0 -\Mat{D}\right)\cdot\delta\tilde{\bm{h}}^{z},
\eeq
and invert the diagonal matrix to get an exact expression for the susceptibility: \beq
\label{inv_EOM}
(\delta\tilde{\bm{s}}^{z})=i\sum_m \Mat{U}\cdot\frac{\bm{\hat{e}}_m\otimes\bm{\hat{e}}_{m}^{T}}{\omega+i\gamma_m}\cdot\Mat{U}^{-1}\cdot\left(\Mat{\Gamma}\cdot\Mat{\tilde{\chi}}_0 -\Mat{D}\right)\cdot \delta\tilde{\bm{h}}^{z}.
\eeq
Transformation (\ref{trans_P}) implies that column vector $\bm{v}_{m}=\bm{{\sf U}} \cdot \bm{\hat{e}}_{m}$ is the right eigenvector of $\Mat{P}$ associated to the eigenvalue $-i\gamma_{m}$. Similarly, the line vector $\bm{v}_{m}^{-1} = \bm{\hat{e}}_{m}^{T} \cdot \bm{{\sf U}}^{-1}$ is the left eigenvector of $\Mat{P}$ associated to the same eigenvalue. We have $\bm{v}^{-1}_m\cdot \bm{v}_{m'}=\delta_{mm'}$, but the set $\{\bm{v}_{m}\}$ is not mutually orthogonal because $[\Mat{P},\Mat{P}^{\dag}]\neq \Mat{0}$ ($\Mat{P}$ is not normal).

Using these results, Eqs.~(\ref{zEOM})~and~(\ref{inv_EOM}) imply the following exact expression for the isotropic dynamical susceptibility:
\begin{subequations}
\begin{eqnarray}
   \bm{{\sf \tilde{\chi}}}^{ab}(\omega) &=& i\left(\omega\Mat{I}-\Mat{P}\right)^{-1}\cdot \left(\Mat{\Gamma}\cdot\Mat{\tilde{\chi}}_0 -\Mat{D}\right)\delta_{ab}\label{dynamical_susceptibility_pre}\\
   &=& i \sum_{m} \frac{\bm{v}_{m} \otimes \left[\bm{v}_{m}^{-1} \cdot \left(\Mat{\Gamma}\cdot\Mat{\tilde{\chi}}_0 -\Mat{D}\right) \right]}{\omega +i\gamma_m}\delta_{ab},
   \label{dynamical_susceptibility}
\end{eqnarray} 
\end{subequations}
valid for $T>T_{c}^{{\rm mag}}$ with $a,b=x,y,z$.
The dynamical susceptibility has poles at the paramagnon frequencies $\omega=-i\gamma_m$, and these contribute to dissipation and noise. The paramagnons are said to be purely dissipative because $\gamma_m$ is real, leading to $e^{-i\omega t}=e^{-\gamma_m t}$ for the decay of the spin excitations. For $T>T_{c}^{{\rm mag}}$ we have $\gamma_{m}\geq 0$ because the paramagnetic phase is stable.

%The fact that $D_{ij}$ conserves spin may be expressed more generally as 

Conservation of total spin follows from $\bm{1}^{T}\cdot \Mat{D}=\bm{0}^{T}$, where  $\bm{1}$ is the column vector with all $N$ components equal to $1$. 
%When in addition $\Mat{\Gamma}=\Mat{0}$, the spin conservation law holds and this has additional implications. 
When in addition $\Mat{\Gamma}=\Mat{0}$ all contributions to the paramagnon matrix (\ref{mat_P}) have $\Mat{D}$ on the left, so it follows that the $m=0$ mode defined by $\bm{\hat{v}}^{-1}_{0}=\bm{1}^{T}/\sqrt{N}$ is a left eigenvector of $\Mat{P}$ associated to $\gamma_{0}=0$. This is true for general $J_{ij}$. Because $\bm{v}_{0}^{-1}\cdot\Mat{D}=\bm{0}^{T}$, $m=0$ does not contribute to the sum in the dynamical susceptibility Eq.~(\ref{dynamical_susceptibility}). This occurs as a direct consequence of the conservation law $\frac{d}{d t} (\bm{1}^{T}\cdot\delta \bm{s}^z)=\bm{0}$, so we say $m=0$ is the nondissipative Goldstone paramagnon.

At high frequency $\omega\gg {\rm Max}_{m}\{\gamma_{m}\}$, Eq.~(\ref{dynamical_susceptibility}) leads to
\beq
   \label{HighwlimitChi}
   \bm{{\sf \tilde{\chi}}}^{ab}(\omega) \approx \frac{i \left(\Mat{\Gamma}\cdot\Mat{\tilde{\chi}}_0-\Mat{D}\right)}{\omega}\delta_{ab},
\eeq
because $\sum_m \bm{v}_{m} \otimes \bm{v}_{m}^{-1} =\Mat{I}$. 
As Eq.~(\ref{eom}) has an upper frequency cut-off, Eq.~(\ref{HighwlimitChi}) should be taken as an upper bound on the modulus of the susceptibility. 

Now consider the opposite limit, $\omega\rightarrow 0$.
Since $\Mat{\tilde{\chi}}^{ab}(0)\equiv\Mat{\tilde{\chi}}_0\delta_{ab}$, 
setting $\omega= 0$ in  Eq.~(\ref{dynamical_susceptibility_pre}) leads to $\left(\Mat{P}+i\Mat{\Gamma}\right)\cdot \Mat{\tilde{\chi}}_0=i\Mat{D}$, and using Eq.~(\ref{mat_P}) we get
\begin{equation}
    \Mat{\tilde{\chi}}^{ab}(\omega=0)=\Mat{\tilde{\chi}}_0\delta_{ab}=\left(4k_BT \Mat{I}- \Mat{J}\right)^{-1}\delta_{ab}.
\label{T_CW_general}
\end{equation}
This is the generalized Curie-Weiss susceptibility for a nonhomogeneous spin system (valid for $T>T_{c}^{{\rm mag}}$). 

Exact analytical results can be obtained for the special case of a translation-invariant spin system. If the system is close to being translation-invariant, e.g. only a few vacancies are present so that $\sigma=\langle x_i\rangle_i \lesssim 1$, and $\Gamma_i$ does not depend appreciably on $i$, a homogeneous approximation (HA) can be proposed. The HA replaces $x_i$, $\Gamma_i$, $J_{ij}$, and $D_{ij}$ by their spatial averages $\sigma$, $\bar{\Gamma}$, $\bar{J}_{ij}$, and $\bar{D}_{ij}$ (the latter two depending only on $\bm{R}_j-\bm{R}_i$), making the problem analytically solvable.  In this case Appendix~\ref{app_homo_approx} shows that the exact susceptibility can be obtained by spatial Fourier transformation, 
\begin{equation}
    \tilde{\chi}^{ab}(\bm{q},\omega)=\frac{i\left(\bar{\Gamma}\tilde{\chi}(\bm{q},0)-\tilde{D}(\bm{q}) \right)\delta_{ab}}{\omega+i\left\{-\tilde{D}(\bm{q})\left[4k_BT-\tilde{J}(\bm{q})\right]+\bar{\Gamma}\right\}},
\label{chizeroq}
\end{equation}
where 
$\tilde{D}(\bm{q})=\sum_{\bm{v}}\bar{D}_{i,i+\bm{v}}e^{-i\bm{q}\cdot\bm{v}}$ and $\tilde{J}(\bm{q})=\sum_{\bm{v}}\bar{J}_{i,i+\bm{v}}e^{-i\bm{q}\cdot\bm{v}}$. The paramagnon modes are labelled by $m=\bm{q}\in 1^{{\rm st}}$~Brillouin zone, each with frequency eigenvalue
\begin{equation}
    \gamma_{m}=\gamma_{\bm{q}}=-\tilde{D}(\bm{q}) \left[ 4 k_B T - \tilde{J}(\bm{q}) \right] + \bar{\Gamma},
\label{gamma_m_TI}
\end{equation}
and associated right and left eigenvectors $\bm{v}_{m}=\bm{e}_{\bm{q}}$, $\bm{v}_{m}^{-1}=\bm{e}_{\bm{q}}^{\dag}/N_s$, respectively, where $\bm{e}^{\dag}_{\bm{q}}=\left(e^{-i \bm{q} \cdot \bm{R}_0 }, \ldots , e^{-i \bm{q} \cdot \bm{R}_{N-1} }\right)$, and $N_s$ is the number of occupied sites forming a translation-invariant lattice. 

When $\omega\rightarrow 0$, Eq.~(\ref{chizeroq}) leads to 
\begin{equation}
    \tilde{\chi}^{ab}(\bm{q},0)=\frac{1}{4k_B}\frac{1}{T-T_{{\rm CW}}(\bm{q})}\delta_{ab}, 
\label{TCW_q}
\end{equation}
where $T_{{\rm CW}}(\bm{q})=\tilde{J}(\bm{q})/(4k_B)$ is the Curie-Weiss temperature in Fourier space.  

Without translation invariance, e.g. in the presence of spin clusters, Eq.~(\ref{T_CW_general}) shows that the $\omega=0$ susceptibility may have several different temperature poles $T_{{\rm CW}}(0)$, each associated with different clusters having different $J_{ij}$ or number of neighbours. However, $T_{{\rm CW}}$ only depends on spin-spin interaction, it does not depend on relaxation parameters $\Gamma_i$. 

\section{Flux noise and paramagnon density \label{section:noise_density}}

The explicit expression for flux noise is obtained by plugging Eq.~(\ref{dynamical_susceptibility}) into Eq.~(\ref{eq_FDT}) and using Eq.~(\ref{eq_fluxnoise})
 \beq \label{S_Phi}
    \tilde{S}_{\Phi}(\omega) = \frac{2 \hbar \omega}{1 - e^{-\frac{\hbar \omega}{k_B T}}} \sum_{m,a}  \frac{\bm{F}^{a T}\cdot\bm{v}_{m} \left[\bm{v}_{m}^{-1} \cdot \left(\Mat{\Gamma}\cdot\Mat{\tilde{\chi}}_0 -\Mat{D}\right)\cdot\bm{F}^{a} \right]}{\omega^2 + \gamma_{m}^2},
\eeq
where $\bm{F}^{a T}=\left(F^{a}(\bm{R}_0),\ldots,F^{a}(\bm{R}_{N-1})\right)$ represents the $a$-component of the flux vector for all spins. 
A convenient way to interpret this expression is to write it in terms of a density of Lorentzian contributions
\begin{equation}
\label{sphirho}
    \tilde{S}_{\Phi}(\omega) =\frac{2 \pi\hbar \omega}{1 - e^{-\frac{\hbar \omega}{k_B T}}}  \int_{-\infty}^{\infty} d \gamma \frac{\gamma / \pi}{\omega^2 + \gamma^2} \rho_{\Phi} (\gamma),
\end{equation}
where $\rho_{\Phi} (\gamma)$ is the \textit{paramagnon flux density}, defined as
\beq \label{eq_paramagnonfluxdensity}
    \rho_{\Phi}(\gamma) = \frac{1}{\gamma}   \sum_{m, a} (\bm{F}^{a T} \cdot \bm{v}_{m})
    \left[\bm{v}_{m}^{-1} \cdot \left(\Mat{\Gamma}\cdot\Mat{\tilde{\chi}}_0 -\Mat{D}\right)\cdot \bm{F}^{a} \right]    
    \delta(\gamma - \gamma_{m}),
\eeq
where $\delta(x)$ is the Dirac delta function. 

When the spin system is translation-invariant, the paramagnon flux density is given by 
\beq
\rho_{\Phi}(\gamma)=\frac{1}{N_s\gamma} \sum_{\bm{q}}\left|\tilde{\bm{F}}(\bm{q})\right|^{2}\left[\bar{\Gamma}\tilde{\chi}(\bm{q},0)-\tilde{D}(\bm{q})\right]\delta(\gamma-\gamma_{\bm{q}}),
\label{paramagnon_flux_density_q}
\eeq
where $\tilde{\bm{F}}(\bm{q})=\sum_{j}\bm{F}_j e^{-i\bm{q}\cdot\bm{R}_j}$ is the flux vector in Fourier space. 

Finally, in the high frequency limit $\omega\gg {\rm Max}_{m}\{\gamma_{m}\}$ Eq.~(\ref{S_Phi}) implies 
\beq
\label{s_Phi_high_omega}
    \tilde{S}_{\Phi}(\omega) = \frac{2 \hbar}{1 - e^{-\frac{\hbar \omega}{k_B T}}} \frac{1}{\omega}\sum_{a}  \bm{F}^{a T}\cdot \left(\Mat{\Gamma}\cdot\Mat{\tilde{\chi}}_0 -\Mat{D}\right)\cdot\bm{F}^{a}.
\eeq
Since Eq.~(\ref{eom}) has an upper frequency cut-off $\Omega_c$, the actual $\tilde{S}_{\Phi}(\omega)$ is expected to be less than Eq.~(\ref{s_Phi_high_omega}) at $\omega>\Omega_c$. In this regime Eq.~(\ref{s_Phi_high_omega}) provides 
an upper bound on flux noise. 

%The next section proposes an explicit form for $D_{ij}$ that leads to spin diffusion. 

\section{Paramagnon flux density in the presence of spin diffusion from an infinite plane of spins\label{sec:infiniteplane}}

Before we display numerical calculations with our explicit expression for $D_{ij}$ shown in Eq.~(\ref{diff_operatorDij}), it is of value to consider a simplified model for  $D_{ij}$ based on the third-principles theory. 
%At this point we consider the first model for $D_{ij}$ and $\rho_{\Phi}(\gamma)$. 
Assume the spin system is an infinite square lattice with no vacancies (homogeneous). Moreover, assume the superconducting wire is also infinite with flux vector given by the ``edge model''shown in Fig.~\ref{FigFlux}:
\beq \label{eq_fluxvector}
\bm{F(\bm{r})} = F_0 (\delta_{x, -W/2} - \delta_{x, W/2}) \hat{\bm{z}},
\eeq
where $W$ is the wire width and $\delta_{x, \pm W/2}$ are Kronecker delta functions. Equation~(\ref{eq_fluxvector}) gives a good description of thin-film wires where it is shown that $\bm{F}_i$ is sharply peaked at the wire edges \cite{LaForest2015}. 

The simplest model for $D_{ij}$ is to emulate the third-principles theory. To do this, assume the phenomenological paramagnon relaxation rates are given by $\gamma_{\bm{q}}=D(T) q^2 + \bar{\Gamma}$, with $D(T)$ the spin diffusion constant. 
Inspection of Eq.~(\ref{gamma_m_TI}) shows that this $\gamma_{\bm{q}}$ is obtained by a choice of $D_{ij}$ that has the following Fourier representation at low $q$:
\beq
\tilde{D}(\bm{q})=-\frac{D(T)q^2}{4\left(k_BT-J\right)}.
\eeq
Plug these into Eq.~(\ref{paramagnon_flux_density_q}) along with $\tilde{\chi}(\bm{q},0)\approx 1/[4(k_BT-J)]$ and $|\tilde{\bm{F}}(\bm{q})|^{2}=4F_{0}^{2}N_{sy}^{2}\sin^{2}{(q_xW/2)}\delta_{q_y,0}$ to get 
\begin{widetext}
\begin{eqnarray}
\rho_{\Phi}(\gamma)&=&\frac{4F_{0}^{2}N_{sy}^{2}}{N_s\gamma}
%\sum_{q_x\in {\rm 1}^{st}~{\rm BZ}} 
\int_{-\pi/a_0}^{\pi/a_0}\frac{dq}{\frac{2\pi}{N_{sx}a_0}}
\sin^{2}{\left(\frac{q_xW}{2}\right)}\frac{\bar{\Gamma}+Dq^2}{4(k_BT-J)}
\delta\left(\gamma-Dq^2-\bar{\Gamma}\right)\nonumber\\
&=& \frac{F_{0}^{2}a_0N_{sy}}{2\pi (k_BT-J)\sqrt{D}}
\sin^{2}{\left[\sqrt{\frac{(\gamma-\gamma_{{\rm min}})W^{2}}{4D}}\right]}
\frac{\theta(\gamma_{{\rm max}}-\gamma)\theta(\gamma-\gamma_{{\rm min}})}{\left(\gamma-\gamma_{{\rm min}}\right)^{1/2}},
\label{rhophiinfinite}
\end{eqnarray}
\end{widetext}
where $\theta(x)$ is the Heaviside step function, with $\gamma_{{\rm min}}=\bar{\Gamma}$ and $\gamma_{{\rm max}}=D(\pi/a_0)^{2}+\bar{\Gamma}$ defining the region where $\rho_{\Phi}(\gamma)$ is nonzero. The sine squared represents interference between the two edges of the wire; in most cases this averages out to $1/2$ either because of small fluctuations in wire shape or frequency resolution during integration over $\gamma$. Apart from this, $\rho_{\Phi}(\gamma)$ follows a power law in frequency
\begin{equation}
    \rho_{\Phi}(\gamma)=\frac{C(T)\theta(\gamma_{{\rm max}}-\gamma)\theta(\gamma-\gamma_{{\rm min}})}{\gamma^{\alpha}},
\label{rhoPhi_power_law}
\end{equation}
with exponent $\alpha=1/2$ and amplitude $C(T)$. 

Plug Eq.~(\ref{rhoPhi_power_law}) into Eq.~(\ref{sphirho}) to get 
\begin{equation}
\label{sphiexp}
    \tilde{S}_{\Phi}(\omega)=\frac{2\hbar\omega}{1-e^{-\frac{\hbar\omega}{k_BT}}}
    C(T)\frac{b_{\alpha}(\omega)}{\omega^{\alpha}}, 
\end{equation}
where 
\begin{equation}
    b_{\alpha}(\omega)=\int_{\gamma_{{\rm min}}/\omega}^{\gamma_{{\rm max}}/\omega}dx \frac{x^{1-\alpha}}{1+x^2}.
\end{equation}
Thus, when $\gamma_{{\rm min}}\ll \omega\ll \gamma_{{\rm max}}$, $b_{\alpha}(\omega)\approx \pi/[2\sin{(\pi\alpha/2)}]$ and the flux noise scales as a power law in frequency with the same exponent $\alpha$, $\tilde{S}_{\Phi}(\omega)\propto 1/\omega^{\alpha}$.

We emphasize that the $\alpha=1/2$ obtained in this section is a direct consequence of assuming long-wavelength diffusion in an infinite and homogeneous (spin density $\sigma=1$) lattice of spins. The next section 
shows explicit numerical calculations of 
%sets up a method to investigate 
the impact of spatial confinement and disorder using Eq.~(\ref{diff_operatorDij}) for $D_{ij}$.

\begin{figure*}
\begin{center}
\subfloat[]{\includegraphics[width=.49\linewidth]{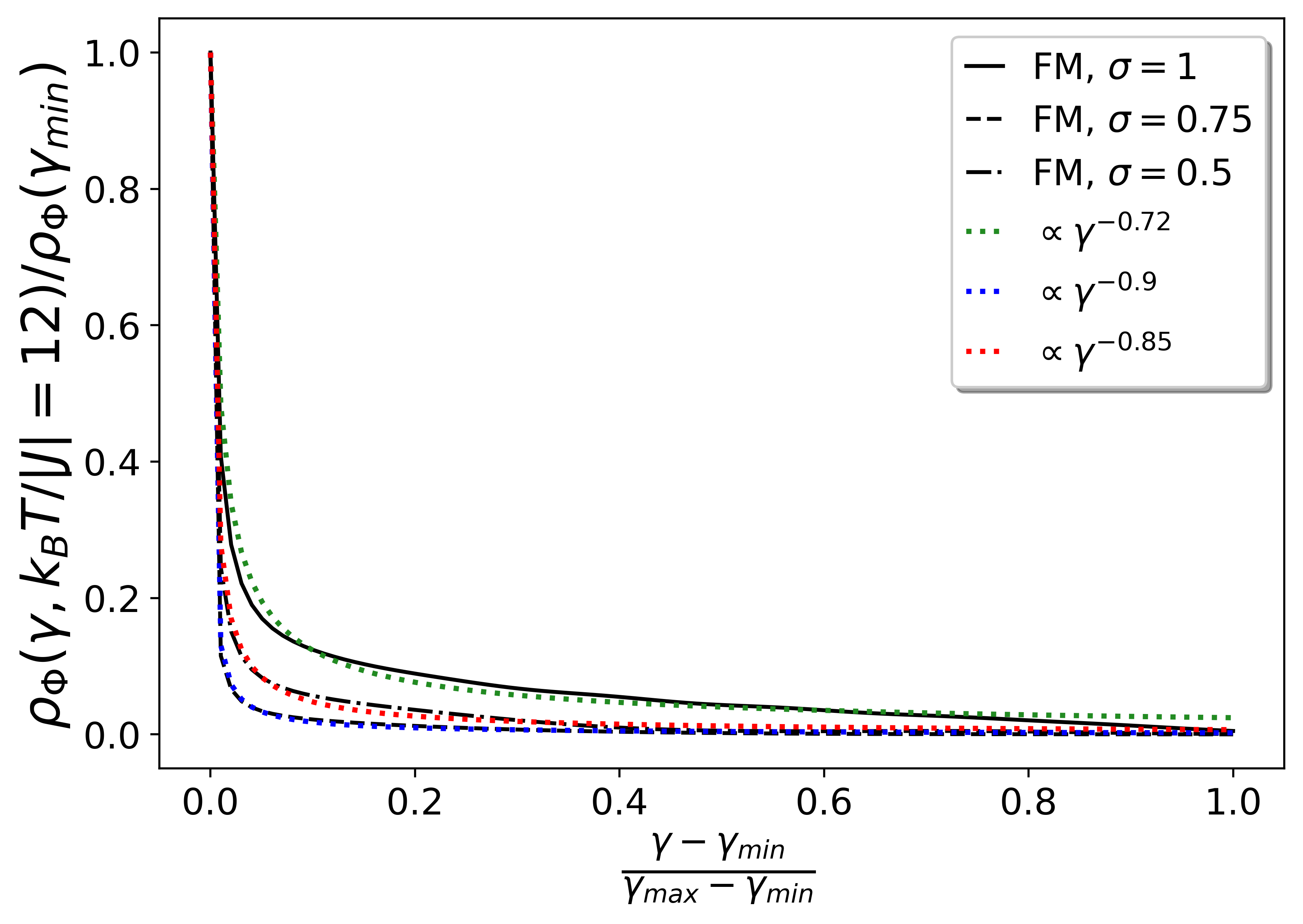}
\label{fig:RhoPhiFM}}
%\hspace{1.5cm}
\subfloat[]{\includegraphics[width=.49\linewidth]{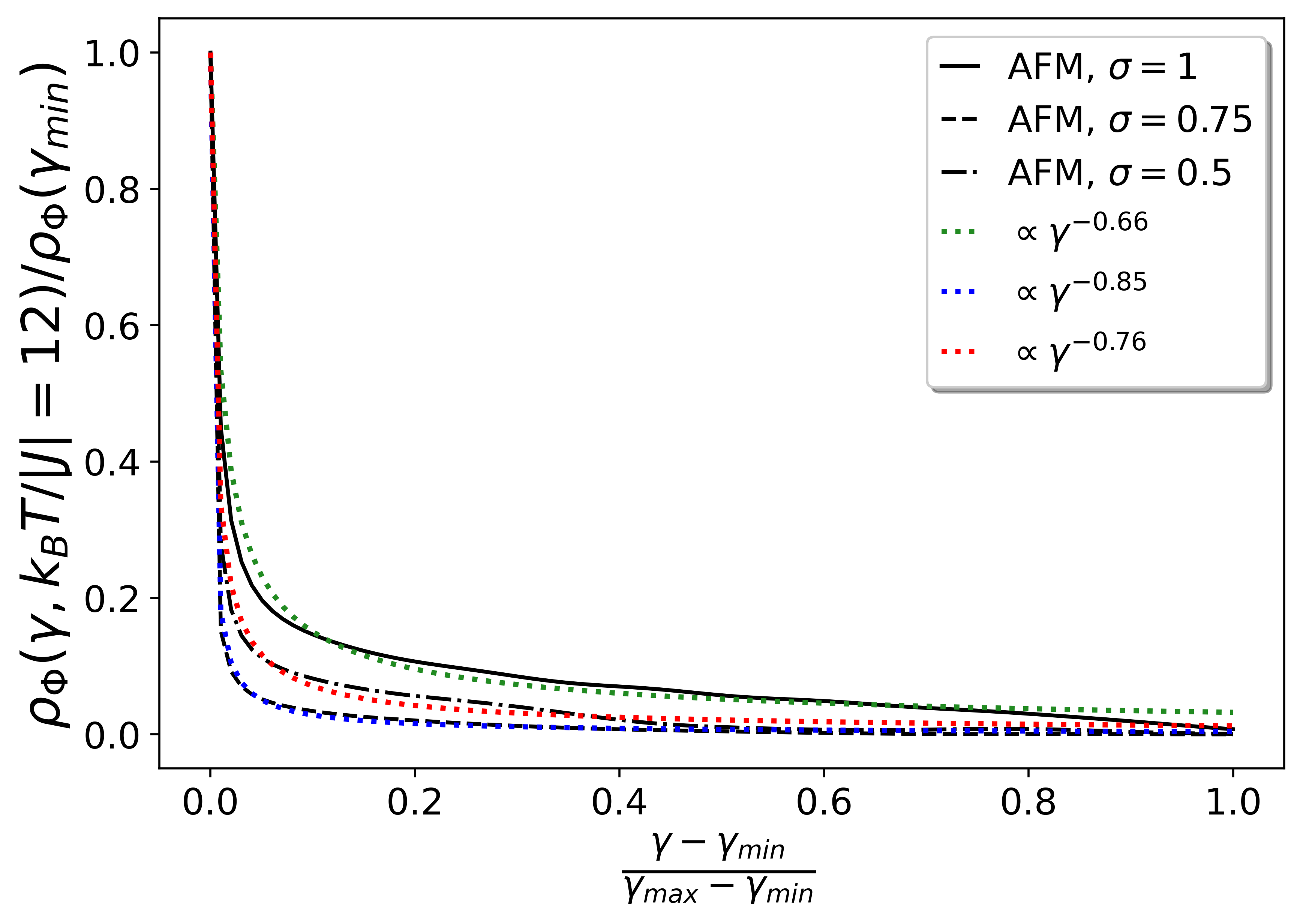}\label{fig:RhoPhiAFM}}
\end{center}
\caption{Explicit calculations of paramagnon flux density $\rho_{\Phi}(\gamma)$ for a confined system of spins ($20 \times 20 $ virtual lattice with open b.c. along $x$), using the edge model for the flux vector (Eq.~(\ref{eq_fluxvector}) and Fig.~\ref{FigFlux}). Results are shown for $k_BT/|J|=12$ and $\Gamma_i=0$ for (a) $J>0$ (FM) and (b) $J<0$ (AFM), for spin densities $\sigma=1, 0.75, 0.5$. The cases with $\sigma<1$ were averaged over 512 random instances, each containing a different distribution of vacancies in the virtual lattice with the same $N_s= \sigma N$. These plots demonstrate that $\rho_{\Phi}(\gamma)$ remains a power law in $\gamma$, even in the presence of confinement (open b.c.) and disorder ($\sigma<1$). However, the exponent $\alpha>1/2$, in contrast to $\alpha=1/2$ in the absence of confinement and disorder.\label{fig:RhoPhi}}
\end{figure*}

\section{Numerical evaluation of the paramagnon flux density: Heisenberg model in the 2d square lattice with a random distribution of vacancies\label{sec:numericalHeisenberg}}

For numerical calculations with our proposed $D_{ij}$ in Eq.~(\ref{diff_operatorDij}), we consider a $20\times 20$ ($N=400$) ``virtual'' square lattice. We randomly populate $N_s \leq N$ sites with spins, yielding spin density $\sigma=N_s/N$. The remaining unoccupied virtual sites are called vacancies, see Fig.~\ref{FigFlux}. All calculations below are done with open boundary condition (b.c.) along $x$, and periodic b.c. along $y$, describing spins confined within the region of the SC wire. 

All calculations assume the nearest-neighbour (n.n.) Heisenberg model in a square lattice with $J_{ij} = J\sum_{\bm{v}}\delta_{i,j+\bm{v}}$ for $\bm{v}=\pm a_0 \bm{\hat{x}},\pm a_0\bm{\hat{y}}$. Divide matrix~(\ref{mat_P}) by $d_0(T)|J|/\hbar$, so that the eigenvalues $\gamma_m$ are expressed in units of $d_0(T)|J|/\hbar$. That way the problem is now specified by $k_BT/|J|$, $\hbar\Gamma_i/(d_0(T)|J|)$, and  $\hbar\omega/(d_0(T)|J|)$. As a result, when $\Gamma_i=0$, the calculated exponent $\alpha$ is independent of the choice of $d_0(T)$ (although $\gamma_{{\rm min}},\gamma_{{\rm max}}$ do depend on $d_0(T)$). 

Figure~\ref{fig:RhoPhi} shows explicit calculations of $\rho_{\Phi}(\gamma)$ using Eq.~(\ref{eq_paramagnonfluxdensity}) with numerical calculations of the eigenvalues $\gamma_m$ of $\Mat{P}$, with the $\delta(x)$ function approximated by a Gaussian with standard deviation $0.1\gamma_{max}$. The flux vector was given by the ``edge model'', see Eq.~(\ref{eq_fluxvector})  and Fig.~\ref{FigFlux}. Results for $k_BT/|J|=12$ and $\Gamma_i=0$ are shown for both $J>0$ (FM) and $J<0$ (AFM), using spin densities $\sigma=1, 0.75, 0.5$. The latter two are averaged over 512 random instances, each containing a different distribution of vacancies in the virtual lattice with the same $N_s= \sigma N$. These results demonstrate that $\rho_{\Phi}(\gamma)$ remains a power law in $\gamma$, even in the presence of confinement (open b.c.) and spatial disorder ($\sigma<1$). However, the corresponding exponent $\alpha>1/2$, contrasting to the case without confinement/disorder (Section~\ref{sec:infiniteplane}). 

\begin{figure*}
\begin{center}
\subfloat[]{\includegraphics[width=.49\linewidth]{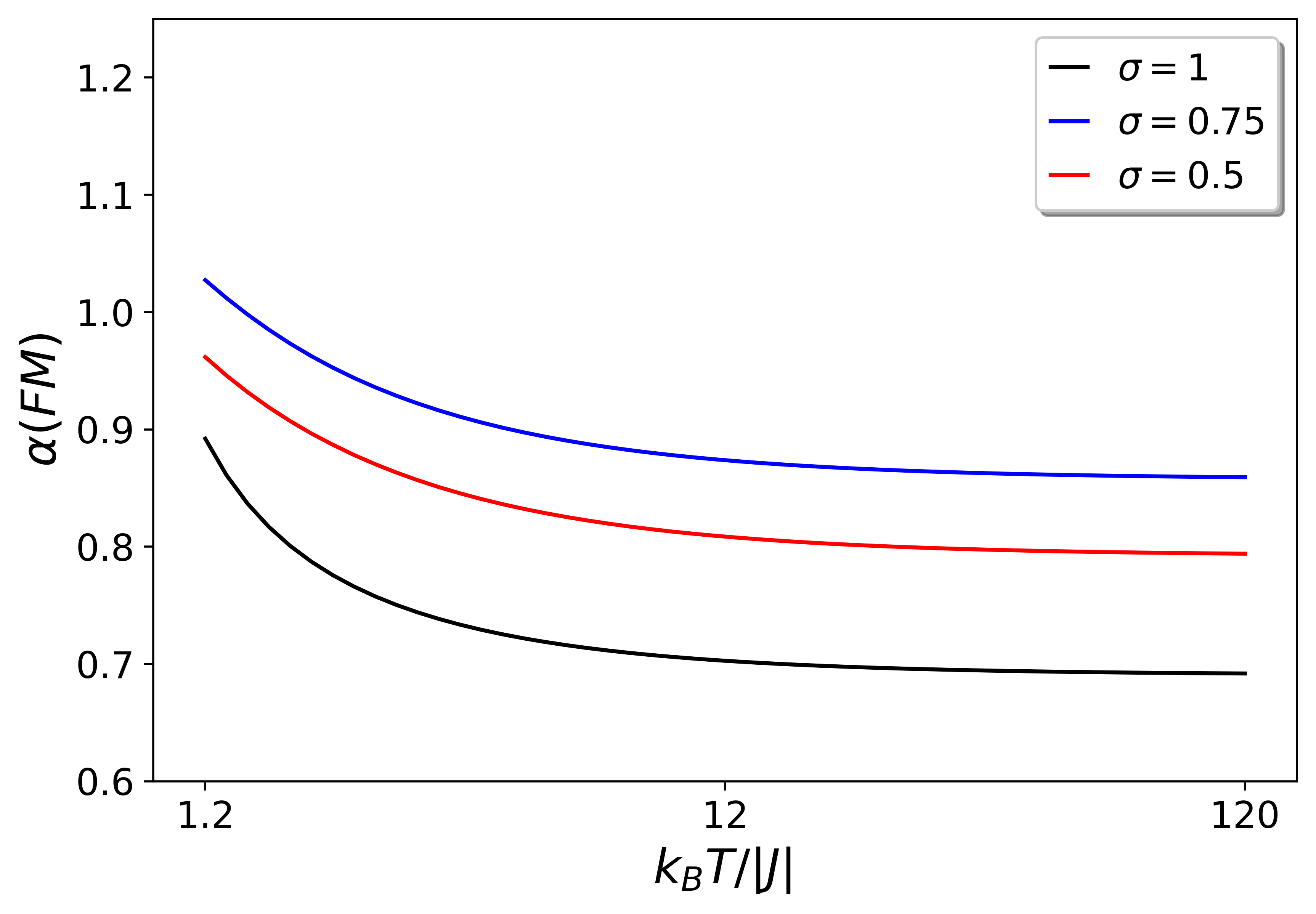}
\label{fig:Alpha_FM}}
\subfloat[]{\includegraphics[width=.49\linewidth]{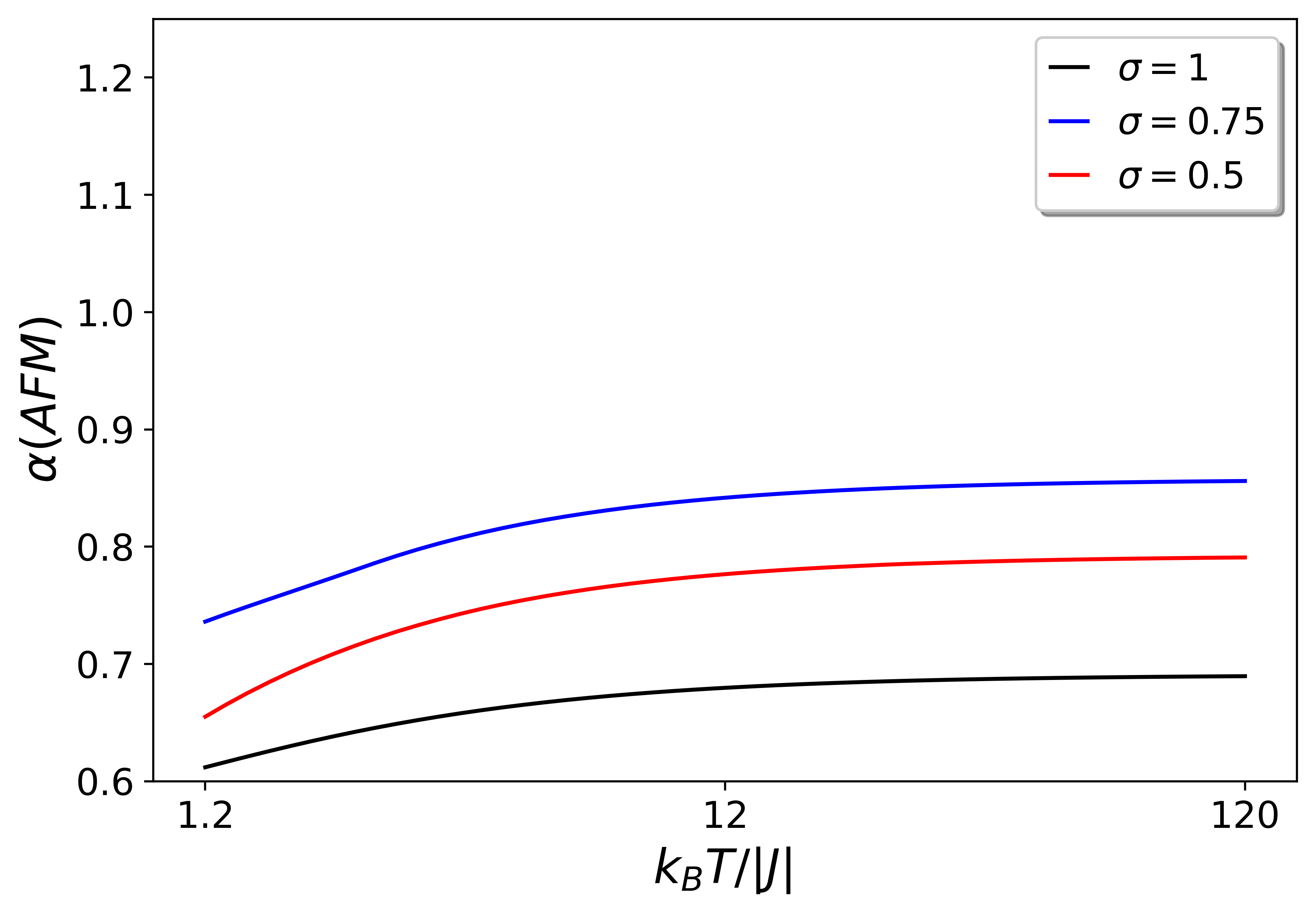}\label{fig:Alpha_AFM}}
\end{center}
\caption{Explicit calculations of flux noise frequency exponent $\alpha$ as a function of temperature for $\Gamma_i=0$ in the n.n. Heisenberg model with confinement and disorder, with all other parameters as in Fig.~\ref{fig:RhoPhi}. (a) FM case with $J>0$, (b) AFM case with $J<0$. Note how $\alpha$ decreases (increases) with $T$ for the FM (AFM) cases. In all cases $\alpha>1/2$ demonstrating the relevance of confinement and disorder. The dependence of $\alpha$ on $\sigma$ is nonmonotonic.\label{fig:Alphavac}}
\end{figure*}

Figure~\ref{fig:Alphavac} shows explicit calculations of the frequency exponent $\alpha$ appearing in both $\rho_{\Phi}(\gamma)$ and $\tilde{S}_{\Phi}(\omega)$ as a function of $T$ and $\sigma$, with all other parameters like in Fig.~\ref{fig:RhoPhi}. It is seen that $\alpha$ \emph{decreases} with $T$ for the FM model, and has the opposite behavior for the AFM model.  At high $T$, both FM/AFM models lead to $\alpha\approx 0.7$ for $\sigma=1$; this demonstrates the importance of confinement. As $\sigma$ decreases from $1$, $\alpha$ further deviates from its infinite/homogeneous value of $1/2$. This demonstrates the impact of disorder. Note how the dependence of $\alpha$ on $\sigma$ is nonmonotonic.

Interestingly, several experiments with niobium devices measure $\alpha=0.7$ \cite{Wellstood87, Zaborniak2021}. 

\begin{figure*}
\begin{center}
\subfloat[]{\includegraphics[width=.49\linewidth]{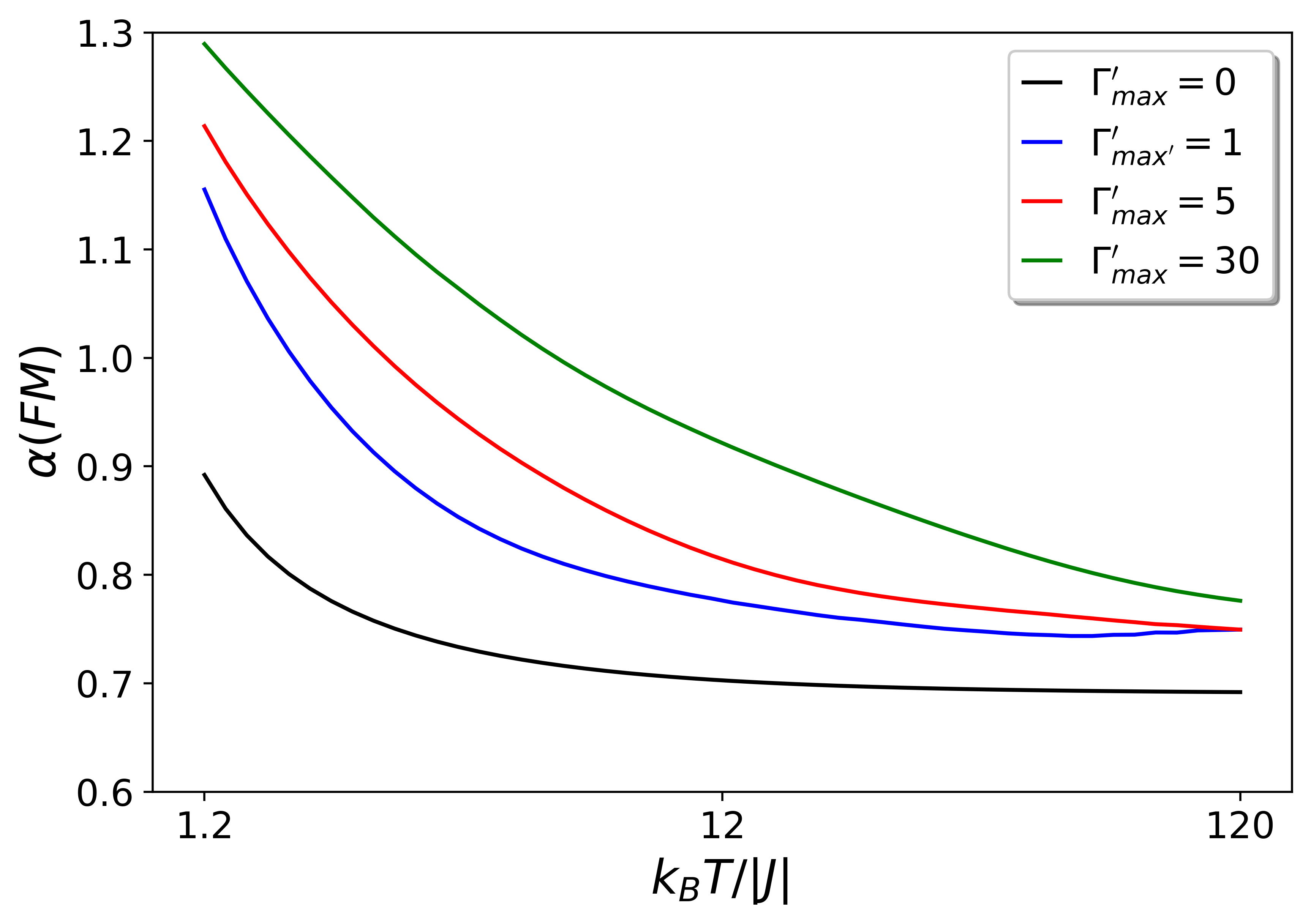}
\label{fig:Alphadiff_FM}}
\subfloat[]{\includegraphics[width=.49\linewidth]{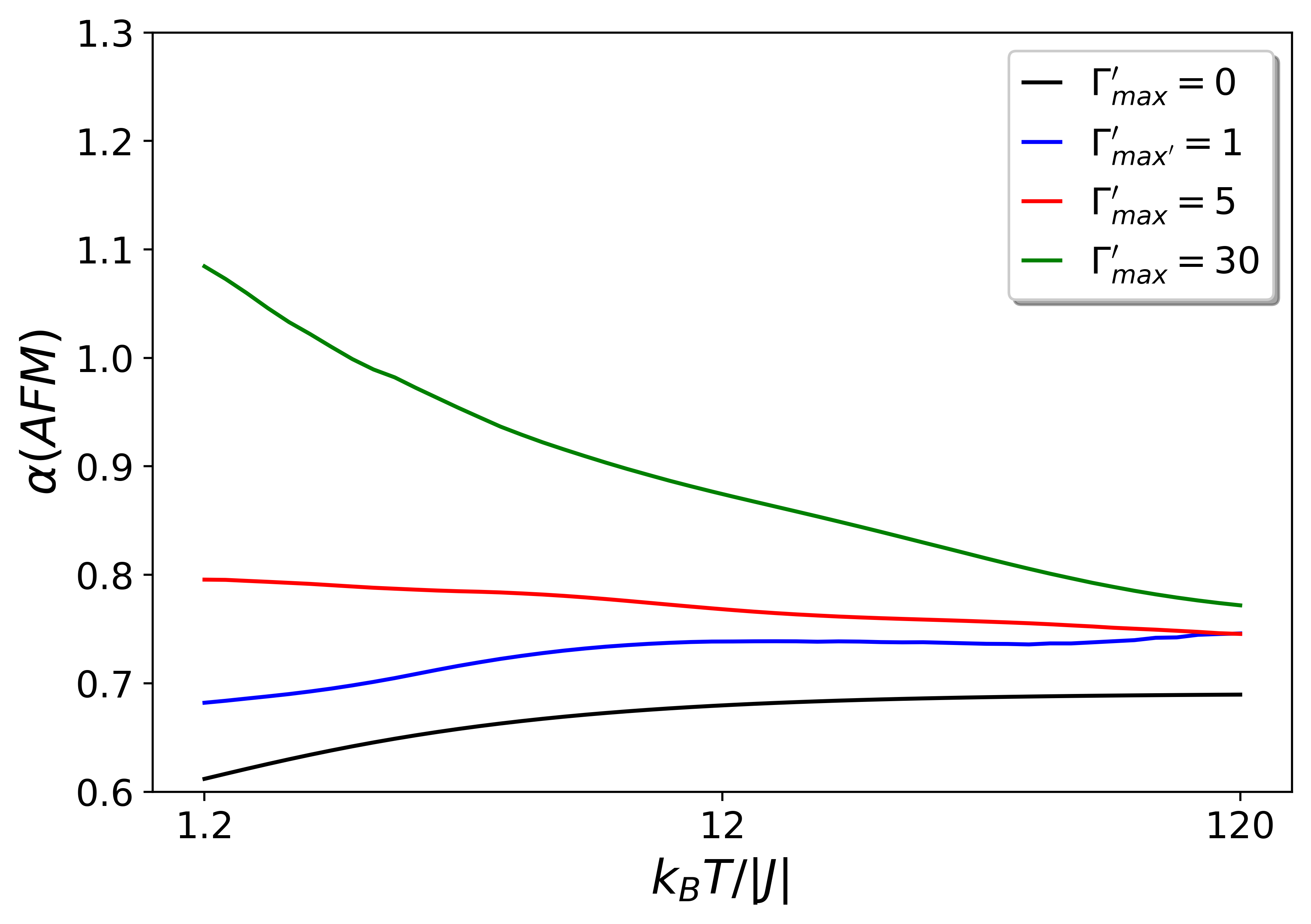}\label{fig:Alphadiff_AFM}}
\end{center}
\caption{Noise frequency exponent $\alpha$ in the presence of both spin-spin interaction $J$ and $\Gamma_i=\Gamma_{{\rm max}}\exp{\left(-\lambda_{{\rm max}}r_i\right)}$, with $r_i\in [0,1]$ a random number for each site $i$. The spin density is $\sigma=1$ for all calculations (no vacancies), so the only source of disorder is the variations in $\Gamma_i$; all other parameters are the same as in Fig.~\ref{fig:Alphavac}. 
The parameters $\Gamma'_{{\rm max}}$ are renormalized relaxation cut-offs, assumed to be constant (see text). Results show that the introduction of $\Gamma_i$ to the interacting spin system increases the values of $\alpha$ in the low to intermediate $T$ range, with the impact of $\Gamma_i$ washed out at high $T$. \label{fig:AlphaGamma}}
\end{figure*}

\section{Disorder due to wide distribution of relaxation rates \texorpdfstring{$\Gamma_i$}{} and spin-spin interactions\label{sec:Gamma}}

Another model for disorder is to assume an arbitrary distribution of relaxation rates $\Gamma_i$. When $J_{ij}=D_{ij}=0$, $\rho_{\Phi}(\gamma)$ and $\tilde{S}_{\Phi}(\omega)$ can be computed exactly. The eigenvalues of $\Mat{P}$ are simply $\gamma_i=\Gamma_i$, and the eigenvectors are the unit column vectors $\hat{\bm{e}}_i$. Using Eq.~(\ref{eq_paramagnonfluxdensity}) the paramagnon flux density becomes
\begin{eqnarray}
    \rho_{\Phi}(\gamma)&=&\frac{1}{\gamma}\sum_i \left|\bm{F}_{i}\right|^{2}\Gamma_i \tilde{\chi}_0 \delta\left(\gamma-\Gamma_i\right)\nonumber\\
    &=& \frac{1}{4k_BT} \sum_i \left|\bm{F}_{i}\right|^{2}\int d\Gamma p(\Gamma) \frac{\Gamma}{\gamma} \delta\left(\gamma-\Gamma\right)\nonumber\\
    &=& \frac{1}{4k_BT} \sum_i \left|\bm{F}_{i}\right|^{2} p(\gamma),
\label{rhophiGamma}
\end{eqnarray}
where in the second line we plugged Eq.~(\ref{T_CW_general}) and took an average using $p(\Gamma)$, the probability density for rates $\Gamma_i$. 
The justification for wide distributions of spin-flip rates $\Gamma_i$ for spin impurities is given in \cite{deSousa2007, Belli2020}. Each spin interacts with one or more amorphous two-level systems (TLSs), leading to the cross-relaxation rate (joint spin flip and TLS switch):
\begin{equation}
    \Gamma_i = \Gamma_{{\rm max}}(T)e^{-\lambda},
\end{equation}
where $\Gamma_{{\rm max}}(T)$ is a cut-off for $\Gamma_i$, and $\lambda$ is a random variable uniformly distributed in the interval $[0,\lambda_{{\rm max}}]$; it models the barrier for TLS switch, see Fig. 3(a) of \cite{deSousa2007}. Such a model is described by the probability density
\begin{equation}
    p(\Gamma)=\frac{1}{\lambda_{{\rm max}}}\frac{1}{\left|\frac{d\Gamma}{d\lambda}\right|}=\frac{1}{\lambda_{{\rm max}}}\frac{1}{\Gamma},
\end{equation}
for $\Gamma_{{\rm min}}(T)<\Gamma<\Gamma_{{\rm max}}(T)$, and zero otherwise, where $\Gamma_{{\rm min}}=\Gamma_{{\rm max}}e^{-\lambda_{{\rm max}}}$. 
Plug this into Eq.~(\ref{rhophiGamma}) and use Eq.~(\ref{sphirho}) to get the flux noise
\begin{equation}
    \tilde{S}_{\Phi}(\omega)=\frac{\pi\hbar\omega}{1-e^{-\frac{\hbar\omega}{k_BT}}}\frac{\sum_i \left|\bm{F}_i\right|^{2}}{4k_BT\lambda_{{\rm max}}}\frac{1}{\omega},
\label{sphi1ofmodel}
\end{equation}
for $\Gamma_{{\rm min}}<\omega<\Gamma_{{\rm max}}$, with constant $\tilde{S}_{\Phi}(\omega)=\tilde{S}_{\Phi}(\Gamma_{{\rm min}})$ for $\omega<\Gamma_{{\rm min}}$, and $\tilde{S}_{\Phi}(\omega)=0$ for $\omega>\Gamma_{{\rm max}}$. From now on we will refer to this model as the \emph{spin $1/f$ model}, since it is the spin equivalent of the well-known $1/f$ noise model in semiconducting devices. 

Now consider the impact of nonzero spin-spin interaction $J_{ij}$ and dissipation $D_{ij}$. Figure~\ref{fig:AlphaGamma} shows numerical calculations using the second-principles theory, for $\sigma=1$ and other parameters as in Fig.~\ref{fig:Alphavac}, plus the spin $1/f$ model with different choices of 
$\Gamma'_{{\rm max}}=\hbar\Gamma_{{\rm max}}(T)/(d_0(T)|J|)$. In Eq.~(\ref{sphi1ofmodel}) we see that the $T$ dependence in $\Gamma_{{\rm max}}$ does not affect the noise provided that $\Gamma_{{\rm min}}<\omega<\Gamma_{{\rm max}}$; this occurs because low frequency noise is independent on the cut-off for $\Gamma_i$. The same argument applies in the presence of $J_{ij}$, provided $\Gamma'_{{\rm max}}$ is a sufficiently large constant in our calculations. 
We simulated the spin $1/f$ model by choosing $\Gamma_i=\Gamma_{{\rm max}}\exp{\left(-\lambda_{{\rm max}}r_i\right)}$ with $\lambda_{{\rm max}}=20$ and $r_i\in [0,1]$ a random number generated for each of the 400 sites.  

The calculations should be compared to the case of $J_{ij}=0$, that has $\alpha=1$ (Eq.~(\ref{sphi1ofmodel})). As shown in Fig.~\ref{fig:AlphaGamma}, the addition of a wide distribution of $\Gamma_i$ to the interacting spin system increases $\alpha$ in the low to intermediate $T$ range. At high $T$, the impact of $\Gamma_i$ is washed out.

\section{Simple expressions for flux noise for comparison to experiments}

The previous sections showed that the paramagnon flux density scales as a power law in $\gamma$, $\rho_{\Phi}(\gamma)=C/\gamma^{\alpha}$, for $\gamma_{{\rm min}}\ll\gamma\ll\gamma_{{\rm max}}$, and $\rho_{\Phi}(\gamma)\approx 0$ outside this range. The exponent $\alpha$ and amplitude $C$ depend on temperature, confinement, and disorder.  

As discussed above Eq.~(\ref{chizeroq}), when $\sigma\lesssim 1$ and $\Gamma_i$ is nearly uniform we can use the homogeneous approximation (HA) described in Appendix~\ref{app_homo_approx} to obtain simpler analytical expressions. 
The HA replaces $x_i$, $\Gamma_i$, $J_{ij}$, and $D_{ij}$ by their spatial averages, $\sigma$, $\bar{\Gamma}$, $\bar{J}_{ij}$, and $\bar{D}_{ij}$, making the problem analytically solvable. 
In particular, $\rho_{\Phi}(\gamma)$ is approximated by the translation-invariant Eq.~(\ref{paramagnon_flux_density_q}). Consider its integral over all $\gamma$:
\begin{eqnarray}
\int d\gamma \rho_{\Phi}(\gamma)&=&\frac{1}{N_s}\sum_{\bm{q}}\left|\tilde{\bm{F}}(\bm{q})\right|^{2}\frac{1}{\gamma_{\bm{q}}}\left[\bar{\Gamma}\tilde{\chi}(\bm{q},0)-\tilde{D}(\bm{q})\right]\nonumber\\
&=& \frac{1}{N_s}\sum_{\bm{q}}\frac{\left|\tilde{\bm{F}}(\bm{q})\right|^{2}}{4k_BT-\tilde{J}(\bm{q})},\nonumber\\
&=& \frac{\sigma \sum_i \left|\bm{F}_i\right|^{2}}{4\left(k_BT-\sigma\bar{J}\right)}, 
\end{eqnarray}
where in the second line we used Eqs.~(\ref{gamma_m_TI})~and~(\ref{TCW_q}),
and in the third line we assumed the typical scale for variations in $\bm{F}(\bm{r})$ is much larger than the lattice spacing $a_0$, so that $\tilde{J}(\bm{q})$ can be approximated by $\tilde{J}(0)=4\sigma \bar{J}$. Now equate this to $\int d\gamma C/\gamma^{\alpha}\approx C\gamma_{{\rm max}}^{1-\alpha}/(1-\alpha)$ (valid for $\alpha<1$) in order to obtain an expression for amplitude $C$; plug this into Eq.~(\ref{sphiexp}) to get 
\begin{equation}
     \tilde{S}_{\Phi}(\omega)=\frac{\hbar\omega}{1-e^{-\frac{\hbar\omega}{k_BT}}}\frac{\sigma (1-\alpha)\sum_i \left|\bm{F}_i\right|^{2}}{2\gamma_{{\rm max}}^{1-\alpha}\left(k_BT-\sigma\bar{J}\right)}
    \frac{b_{\alpha}(\omega)}{|\omega|^{\alpha}}. 
\label{sphi_compexp}
\end{equation}
When $\alpha\lesssim 1$, the temperature dependence of $\gamma_{{\rm max}}\propto d_0(T)T$ is washed out, leading to a simple relation valid for $\hbar\omega \ll k_BT$:
\begin{equation}
    \frac{\tilde{S}_{\Phi}(\omega, T)}{\tilde{S}_{\Phi}(\omega, T\gg \sigma \bar{J}/k_B)}\approx \frac{k_BT}{k_BT - \sigma \bar{J}}.
\label{noiseratioanalytic}
\end{equation}
Note how within HA and for $\alpha\lesssim 1$ this noise amplitude ratio is independent of $d_0(T)$ and other details such as values of $\Gamma_i$.

\begin{figure}
\begin{center}
\includegraphics[width=.99\linewidth]{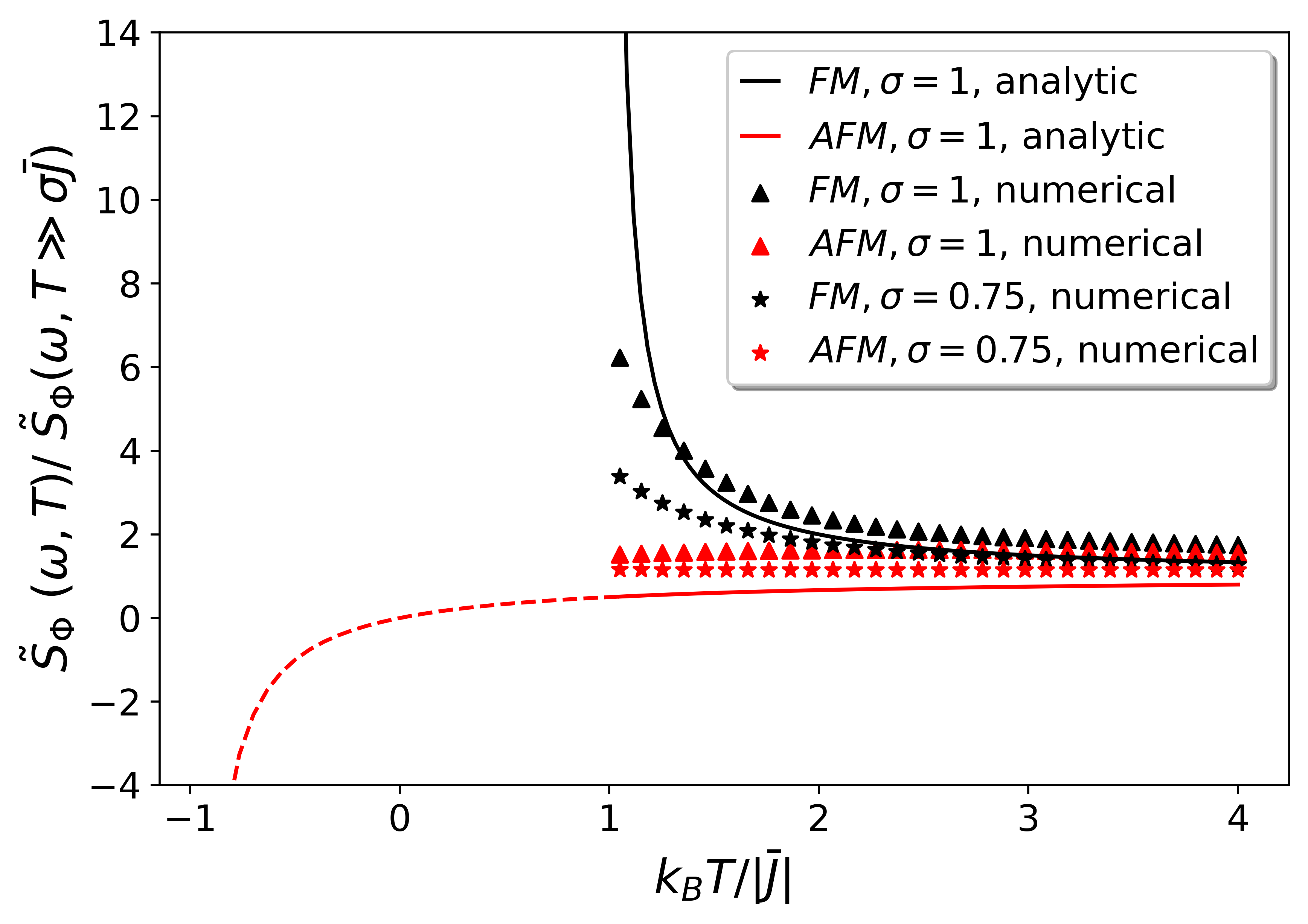}
\end{center}
\caption{Temperature dependence of flux noise amplitude. Due to the formation of magnetized spin clusters, the noise amplitude for the FM model has a singularity at $T=T_{c}^{{\rm mag}}=\sigma|\bar{J}|/k_B$. As a result, the FM $\tilde{S}_{\Phi}(\omega)$ decreases with increasing $T$. For the AFM model ($\bar{J}<0$), $\tilde{S}_{\Phi}(\omega)$ is instead slightly increasing with $T$. The solid curves are based on Eq.~(\ref{noiseratioanalytic}), with dashed ones representing simple  extrapolation to determine $\sigma|\bar{J}|$ for the AFM model. Data points are numerical evaluation with the second principles theory, using $T=10 \bar{J}/k_B$ and $\hbar \omega / |\bar{J}| = 0.1$ as the high $T$ and low $\omega$, respectively, and $d_0(T)=1$, $\Gamma_i=0$.  \label{fig:NoiseAmpRatio}}
\end{figure}

Figure~\ref{fig:NoiseAmpRatio} shows this dependence for the FM ($\bar{J}>0$) and AFM ($\bar{J}<0$) cases, and compares Eq.~(\ref{noiseratioanalytic}) to numerical evaluation with $T=10 \bar{J}/k_B$ and $\hbar \omega / |\bar{J}| = 0.1$ chosen as the high temperature and frequency, respectively, and $d_0(T)=1$, $\Gamma_i=0$. It shows that the amplitude of $\tilde{S}_{\Phi}(\omega, T)$ for a given $\omega$ \emph{decreases} with increasing $T$ for the FM model. For the AFM model it instead \emph{increases} with $T$. Extrapolating to lower temperatures (including $T<0$ for AFM) allows determination of the value of $\sigma\bar{J}$ that models the spins.

The singularity in $\tilde{S}_{\Phi}(\omega)$ for the FM model as $T$ gets close to $T_{c}^{{\rm mag}}=\sigma\bar{J}/k_B$ is due to the formation of clusters of spins with nonzero magnetization (short range order) \cite{Anton2013, LaForest2015}. As seen in Fig.~\ref{fig:NoiseAmpRatio}, numerical results deviate from Eq.~(\ref{noiseratioanalytic}) demonstrating the effect of confinement, disorder and the value of $\alpha$ ($\alpha < 1$ for most T, see Fig. \ref{fig:Alphavac}) on the $T$ dependence of spin clusters. 

Similar considerations apply to the antisymmetric flux noise 
\begin{eqnarray}
    \tilde{S}^{-}_{\Phi}(\omega)&=&\tilde{S}_{\Phi}(\omega)-\tilde{S}_{\Phi}(-\omega)\nonumber\\
    &=& \frac{\sigma (1-\alpha)\sum_i \left|\bm{F}_{i}\right|^{2}}{2 \gamma_{{\rm max}}^{1-\alpha}\left(k_BT -\sigma\bar{J}\right)}\hbar|\omega|^{1-\alpha}b_{\alpha}(\omega). 
\label{sphiminus_highT}
\end{eqnarray}
This is seen to have weak frequency dependence when $\alpha\lesssim 1$. Plotting $1/\tilde{S}^{-}_{\Phi}(\omega)$ as a function of $T$ yields a straight line that extrapolates to zero at $T=\sigma\bar{J}<0$. In Quintana {\it et al.} \cite{Quintana2017} (inset of Fig.~3) this procedure reveals $\sigma\bar{J}=-10$~mK. 

\section{Discussion and conclusions}

Several experiments in SQUID-based devices \cite{Wellstood87, Anton2013, Lanting2014a, Yan2016, Quintana2017, Zaborniak2021} have concluded that flux noise follows the empirical law 
\begin{equation}
    \tilde{S}_{\Phi}(\omega)=\frac{A}{\omega^{\alpha}},
\end{equation}
where amplitude $A$ and exponent $\alpha<1$ are both temperature-dependent. Here we developed a theory of flux noise due to interacting spins that is able to calculate $A$ and $\alpha$ for realistic model impurity spin systems with disorder due to vacancies and wide distributions of spin relaxation rates, and for spins confined in bounded regions such as SC wires. 

To achieve this we needed to develop a method that does not rely on the ``third-principles'' diffusion operator $D\nabla^2$. \Ch{Our ``second-principles'' method instead assumes lattice sites are coupled by a dissipation matrix $D_{ij}$ given by Eq.~(\ref{diff_operatorDij}). We showed that this choice obeys fundamental principles such as total spin conservation and the second law of thermodynamics. Our prescription for $D_{ij}$ depends on spin Hamiltonian parameters such as exchange interaction, establishing a direct connection between flux noise and model spin Hamiltonians}.

%accounts for the spin conservation law exactly, and directly relates flux noise to the parameters of the chosen spin Hamiltonian.
%
%We described results in terms of a general $D_{ij}$, and also proposed a particular form for $D_{ij}$ based on fundamental principles such as the second law of thermodynamics. 
%
%This allowed us to write $D_{ij}$ in terms of the exchange interaction between spins, connecting our theory to model spin Hamiltonians.

A central concept is the interpretation of flux noise in terms of (para)magnon excitations. Flux noise is shown to be directly related to the density of edge (para)magnons, the ones that lead to fluctuations at the superconducting wire edges, where spin flips cause the largest flux changes to the device. While $D_{ij}$ accounts for the interactions between (para)magnons, the rates $\Gamma_i$ describe interactions between spins and other degrees of freedom such as phonons, electron gas excitations, and two-level system defects. 

\Ch{Section~\ref{sec:infiniteplane} shows that choosing $D_{ij}$ consistent with $D\nabla^2$ (third-principles theory) and assuming an infinite, translation-invariant spin system leads to temperature-independent noise exponent $\alpha=1/2$, in contradiction to experiments. In contrast, Sections~\ref{sec:numericalHeisenberg}~and~\ref{sec:Gamma} show that numerical calculations with the ``second principles'' prescription for $D_{ij}$ (Eq.~(\ref{diff_operatorDij})) makes $\alpha$ temperature-dependent and in the range observed in experiments. Both $D_{ij}$ and $\Gamma_{i}$ seem to be required to explain experiments}. 

We can separate experiments in two groups, the ones using niobium and the ones using aluminum devices. Measurements in niobium devices show both amplitude $A$ and exponent $\alpha$ decreasing with increasing $T$ (Fig.~3 of \cite{Anton2013}). According to our Eq.~(\ref{noiseratioanalytic}) and Fig.~\ref{fig:NoiseAmpRatio} this $A(T)$ requires a ferromagnetic model ($J>0$). Our calculated $\alpha(T)$ for $\Gamma_i=0$ ranges from $0.9$ at low $T$ to $0.7$ at high $T$ for a confined system (spins only on top of the wire) with a small number of vacancies ($\sigma\lesssim 1$), but the exponent can approach $0.5$ when the spin system extends beyond the wire (Eq.~(\ref{rhophiinfinite})). In comparison, experimental measurement shows $\alpha(T)$ going from $0.8$ to $0.4$ with increasing $T$ 
\cite{Anton2013}. Therefore, a n.n. Heisenberg model with $J>0$ and $\sigma\lesssim 1$ and low $\Gamma_i\ll J/\hbar$ provides a reasonable model. 

Much less data exists for aluminum devices, but one experiment definitely shows that their $A(T)$ and $\alpha(T)$ are qualitatively different from niobium.  In \cite{Quintana2017}, $A(T)$  is shown to increase with $T$, and based on our Fig.~\ref{fig:NoiseAmpRatio} an antiferromagnetic model ($J<0$) is required. The authors reached the same conclusion by measuring asymmetric noise $\tilde{S}_{\Phi}^{-}(\omega)$ and extrapolating to $T<0$ to get $T_{{\rm CW}}=-10$~mK. According to our interpretation this implies $\sigma \bar{J}/k_B=-10$~mK. The measured exponent $\alpha=0.96-1.05$ can not be explained by our theory with $\Gamma_i=0$ (Fig.~\ref{fig:Alpha_AFM}). However, Fig.~\ref{fig:Alphadiff_AFM} shows that introducing a wide distribution of $\Gamma_i$'s (the spin $1/f$ model) makes $\alpha\approx 1$  at low temperatures. This indicates the necessity of a model with both $\Gamma_i$ and $J$ nonzero in aluminum devices.  
Measurements of $\alpha(T)$ over a wide temperature range are not yet available to confirm this scenario. 

In  addition to low frequency flux noise, experiments also measure Ohmic ($\propto \omega$) \cite{Lanting2011} or super-Ohmic ($\propto \omega^3$) \cite{Yan2016, Quintana2017} flux noise in the GHz range. We now argue that this behaviour can not arise from interacting spins alone. 

Our theory is fundamentally based on the assumption of ``hydrodynamics'', i.e. that spin degrees of freedom can be described by the classical equation of motion (\ref{eom}) \cite{Kadanoff1963}. As a result, it overestimates the noise for $\omega> \Omega_c$, where $\Omega_c$ is a high-frequency cut-off. 
The cut-off $\Omega_c$ can be estimated from exact calculations of the moments of the noise spectrum at $T\rightarrow \infty$. Calculations for the 3d Heisenberg model \cite{DeGennes1958} suggests $\hbar\Omega_c\sim 10 J/\hbar$ for our 2d case. When $\omega >\Omega_c$, $\tilde{S}_{ij}(\omega)$ drops off faster than $1/\omega^2$, so that our  Eq.~(\ref{S_Phi}) becomes an upper bound for flux noise. In Eq.~(\ref{s_Phi_high_omega}) this upper bound was shown to be $\propto 1/\omega$ when quantum noise is included. As the $1/\omega$ upper bound holds for all interacting spin models, this allows us to conclude that the high frequency Ohmic \cite{Lanting2011} or super-Ohmic \cite{Yan2016, Quintana2017} flux noise observed in SQUIDs \emph{can not originate from a model of interacting impurity spins}. A likely source is the normal resistance due to excited quasiparticles, either by thermal or nonequilibrium sources such as cosmic rays \cite{Wilen2021}. 

In conclusion, we developed a ``second principles'' theory of flux noise due to interacting spins that is able to account for the confinement and disorder present in realistic impurity spins systems on superconducting devices. The theoretical framework allows explicit prediction of the amplitude and exponent of flux noise due to different wire geometries and spin disorder scenarios, such as random vacancies and wide distributions of spin-flip rates due to interactions with amorphous TLSs. Comparing numerical results to experiments allowed us to specify different spin Hamiltonians for niobium and aluminum devices. 
Generalizations of the theory that include time dependent external currents and fields can be used to design of optimal control strategies that reduce the impact of flux noise on quantum devices. 

\begin{acknowledgments}
This work was supported by NSERC (Canada) through its Discovery program (Grant number RGPIN-2020-04328). The authors thank M. Amin, R. Harris, P. Kovtun, and T. Lanting for useful discussions. 
\end{acknowledgments}

\appendix 

\section{Exact solution of the homogeneous case and homogeneous approximation \label{app_homo_approx}} 

When the spin system is translation-invariant, the ``second principles'' method yields exact analytic expressions for the spin noise for general $D_{ij}$ using spatial Fourier transforms. This is the case when the system has periodic b.c., and the vacancies are organized in a regular sublattice of the full virtual lattice. The homogeneous case (no vacancies, $\sigma=1$) with periodic b.c. is the most relevant example. 

When the system is not translation invariant, we may take spatial averages over parameters $x_i$, $D_{ij}$, $J_{ij}$, and $\Gamma_i$ in order to force its equation of motion to become exactly solvable. In this case the resulting analytic solution is called homogeneous approximation (HA). 

In the paramagnetic phase ($T>T_{c}^{{\rm mag}}$), the EOM Eq.~(\ref{eq_eom_linear}) becomes
\begin{eqnarray}
    \frac{d}{d t} (\delta s_{i}^{a}) &=& -\sum_j D_{ij} x_j \left[(\delta h_{j}^{a})+\sum_k x_k J_{jk} (\delta s_{k}^{a})\right.\\
    &&\left.-4 k_B T (\delta s_{j}^{a})\right] -\Gamma_i (\delta s_{i}^{a})+\Gamma_i\sum_j\tilde{\chi}_{ij}(0)(\delta h_{j}^{a}),\nonumber
\end{eqnarray}
for $a=x,y,z$.
If this is not translation invariant, replace $x_i, D_{ij}, \Gamma_i$ by their average values:
\begin{subequations}
    \begin{eqnarray}
        \bar{x} &=& \frac{1}{N}\sum_i x_i = \sigma,\\
        \bar{D}_{\bm{v}} &=& \frac{1}{N \sigma^2} \sum_i x_i x_{i+\bm{v}} D_{i,i+\bm{v}},\\
        \bar{J}_{\bm{v}}&=&\frac{1}{N\sigma^2}\sum_{i}x_ix_{i+\bm{v}}J_{i,i+\bm{v}},\label{Javg}\\
        \bar{\Gamma} &=& \frac{1}{N \sigma} \sum_i x_i \Gamma_i,\\
        \bar{\chi}_{\bm{v}}(0) &=& \frac{1}{N \sigma^2} \sum_i \tilde{\chi}_{i,i+\bm{v}}(0).
    \end{eqnarray}
\end{subequations}
Equation~(\ref{eq_eom_linear}) becomes
\begin{eqnarray}
        \frac{d}{d t} (\delta s_{i}^{a}) &=& -\sum_{\bm{v}} \sigma \bar{D}_{\bm{v}} \left[(\delta h_{i+\bm{v}}^{a})+\sum_{\bm{v}'} \sigma \bar{J}_{\bm{v}'} (\delta s_{i+\bm{v}+\bm{v}'}^{a})\right.\\
    &&\left.-4 k_B T (\delta s_{i+\bm{v}}^{a})\right] -\bar{\Gamma} (\delta s_{i}^{a})+\bar{\Gamma}\sum_{\bm{v}}\bar{\chi}_{\bm{v}}(0)(\delta h_{i+\bm{v}}^{a}).\nonumber
\end{eqnarray}
Take the Fourier transform in both sides
\beq
\delta\tilde{s}^{a}_{\bm{q}}(\omega)=\int dt \sum_j e^{-i\left(\bm{q}\cdot\bm{R}_{j}-\omega t\right)}(\delta s_{j}^{a}), 
\eeq
to obtain the dynamical susceptibility,
\begin{equation}
    \tilde{\chi}^{ab}(\bm{q},\omega)
      =\frac{\left[\bar{\Gamma}\tilde{\chi}(\bm{q},0)-\tilde{D}(\bm{q})\right]\delta_{ab}}{-i\omega+\left\{-\tilde{D}(\bm{q})\left[4k_BT-\tilde{J}(\bm{q})\right]+\bar{\Gamma}\right\}}, 
\end{equation}
where
\begin{subequations}
\begin{eqnarray}
        \tilde{D}(\bm{q})&=&\sigma\sum_{\bm{v}}\bar{D}_{\bm{v}}e^{-i\bm{q}\cdot\bm{v}},\\
        \tilde{J}(\bm{q})&=&\sigma\sum_{\bm{v}}\bar{J}_{\bm{v}}e^{-i\bm{q}\cdot\bm{v}}.\label{Javg_q}
\end{eqnarray}
\end{subequations}
%Note that $\tilde{D}(\bm{-q})=\tilde{D}(\bm{q}), \tilde{J}(\bm{-q})=\tilde{J}(\bm{q})$. 

This result implies the zero-frequency susceptibility,
\begin{equation}
\label{CurieWeiss}
    \tilde{\chi}^{ab}(\bm{q},\omega=0)=\frac{1}{4k_B}\frac{\delta_{ab}}{T-T_{{\rm CW}}(\bm{q})},
\end{equation}
with Curie-Weiss temperature $T_{{\rm CW}}(\bm{q})= \frac{1}{4k_B}\tilde{J}(\bm{q})$.

Using the fluctuation-dissipation theorem the spin noise in Fourier space becomes exactly equal to 
\begin{eqnarray} \label{eq_spinnoise_zz_homo}
        \tilde{S}^{ab}(\bm{q}, \omega) &=& \frac{2 \hbar \omega}{1 - e^{- \hbar \omega / k_B T}}\\ &&\times\frac{\left[\bar{\Gamma}\tilde{\chi}(\bm{q},0)-\tilde{D}(\bm{q})\right]\delta_{ab}}{\omega^2 + \left\{  -\tilde{D}(\bm{q}) \left[ 4 k_B T - \tilde{J}(\bm{q}) \right] + \bar{\Gamma} \right\}^2}.\nonumber
\end{eqnarray}
In the notation of Eq.~(\ref{dynamical_susceptibility}), the paramagnon modes are labelled by $m=\bm{q}\in 1^{{\rm st}}$~Brillouin zone, each with frequency eigenvalue
\begin{equation}
    \gamma_{m}=\gamma_{\bm{q}}=-\tilde{D}(\bm{q}) \left[ 4 k_B T - \tilde{J}(\bm{q}) \right] + \bar{\Gamma},
\label{gamma_q}
\end{equation}
and right and left eigenvectors $\bm{v}_{m}=\bm{e}_{\bm{q}}$, $\bm{v}_{m}^{-1}=\bm{e}_{\bm{q}}^{\dag}/N_s$, respectively, where $\bm{e}^{\dag}_{\bm{q}}=\left(e^{-i \bm{q} \cdot \bm{R}_0 }, \ldots , e^{-i \bm{q} \cdot \bm{R}_{N-1} }\right)$.

Note that we assumed the presence of $N_s$ $\bm{R}_i$'s forming a translation-invariant lattice, so there are $N_s$ $\bm{q}$'s in the $1^{{\rm st}}$ Brillouin zone. 
As a consequence Eq.~(\ref{eq_spinnoise_zz_homo}) does not have $\sigma$ appearing explicitly in the numerator.

Similar to what was done in Section~\ref{section:noise_density}, we can define a mode density to interpret Eq.~(\ref{eq_spinnoise_zz_homo}),
\begin{eqnarray}
\label{eq_spinnoise_zz}
    \tilde{S}^{ab}(\bm{q}, \omega ) =\frac{2 \pi \hbar \omega}{1 - e^{-\frac{\hbar \omega}{k_B T}}}  \int d \gamma \frac{\gamma / \pi}{\omega^2 + \gamma^2} \rho^{ab} (\gamma, \bm{q}). 
\end{eqnarray}
The $\rho^{ab} (\gamma, \bm{q})$ is called \textit{paramagnon wavevector density} (in contrast to the paramagnon flux density defined by Eq.~(\ref{eq_paramagnonfluxdensity})). It's given by 
\begin{equation}
    \rho^{ab}(\gamma, \bm{q})=\frac{1}{\gamma}\left[\bar{\Gamma}\tilde{\chi}(\bm{q},0)-\tilde{D}(\bm{q})\right]\delta(\gamma-\gamma_{\bm{q}})\delta_{ab}.
    \label{eq_paramagnondensity_homo}
\end{equation}

For the n.n. Heisenberg model in the 2d square lattice we get 
\begin{subequations}
\begin{eqnarray}
    \tilde{D}(\bm{q})&=&-\frac{d_0(T)}{\hbar}\left[\sin^{2}{\left(\frac{q_x a_0}{2}\right)}+\sin^{2}{\left(\frac{q_y a_0}{2}\right)}\right],\\
    \tilde{J}(\bm{q})&=&2\sigma J \left[\cos{(q_xa_0)}+\cos{(q_ya_0)}\right].
\end{eqnarray}
\end{subequations}
We remark that $\tilde{D}(\bm{q})$ does not depend on $\sigma$ because from Eq.~(\ref{eq_jc}) $\bar{J}_c=4\sigma|J|$, so that $\sigma$ cancels out in the definition of $\tilde{D}(\bm{q})$. 
These results lead to the effective diffusion constant in the HA, 
\begin{eqnarray}
\label{d_hom}
    D_{{\rm hom}}(\bm{q})&\equiv& -\frac{\tilde{D}(\bm{q}) \left[ 4 k_B T - \tilde{J}(\bm{q}) \right]}{q^2}\\
    &=& \frac{d_0(T)a_{0}^{2}}{\hbar}\left[k_BT-\sigma J\left(1-\frac{q^2a_{0}^{2}}{4}\right)\right]+{\cal O}(q^4).\nonumber
\end{eqnarray}

Finally, in the HA approximation the flux noise with edge flux vector Eq.~(\ref{eq_fluxvector})
is given by
\beq \label{eq_fluxnoise_homo}
    \tilde{S}_{\Phi} (\omega) = 4 F_{0}^{2}\frac{N_{sy}}{N_{sx}} \sum_q \sin^{2}{\left(\frac{qW}{2}\right)}\tilde{S}^{zz}(q\bm{\hat{x}}, \omega),
\eeq
where $q=\frac{2\pi}{N_{sx} a_0}\left(n-\frac{N_{sx}}{2}\right)$ with $n=0,1,\ldots, N_{sx}-1$, where $N_{s}=N_{sx}N_{sy}$ is the number of occupied sites.  

\bibliography{fluxnoise_Alberto}

\end{document}